\documentclass[a4paper,fleqn,usenatbib]{mnras}

\usepackage{newtxtext,newtxmath}
\usepackage[T1]{fontenc}
\usepackage{ae,aecompl}

\usepackage{graphicx}	
\usepackage{subfig}
\usepackage{amsmath}	
\usepackage{amssymb}	
\usepackage{inputenc}
\usepackage{color}
\title[Simulations of rotating two component jets]{Rotation and toroidal magnetic field effects on the stability of two-component jets}

\author[D. Millas et al.]{
Dimitrios Millas,$^{1}$\thanks{E-mail: dimitrios.millas@kuleuven.be}
Rony Keppens$^{1}$
and Zakaria Meliani$^{2}$
\\
$^{1}$Centre for Mathematical Plasma Astrophysics, Department of Mathematics, KU Leuven, Leuven, Belgium\\
$^{2}$LUTH, Observatoire de Paris-Meudon, Paris, France\\
}

\date{Accepted XXX. Received YYY; in original form ZZZ}

\pubyear{2017}

\begin{document}
\label{firstpage}
\pagerange{\pageref{firstpage}--\pageref{lastpage}}
\maketitle
\begin{abstract}
Several observations of astrophysical jets show evidence of a structure in the direction perpendicular to the jet axis, leading to the development of ``spine \& sheath'' models of jets. Most studies focus on a two-component jet consisting of a highly relativistic inner jet and a slower -- but still relativistic -- outer jet surrounded by an unmagnetized environment. These jets are believed to be susceptible to a relativistic Rayleigh-Taylor-type instability, depending on the effective inertia ratio of the two components. We extend previous studies by taking into account the presence of a non-zero toroidal magnetic field. Different values of magnetization are examined, to detect possible differences in the evolution and stability of the jet. We find that the toroidal field, above a certain level of magnetization $\sigma$, roughly equal to 0.01, can stabilize the jet against the previously mentioned instabilities and that there is a clear trend in the behaviour of the average Lorentz factor and the effective radius of the jet when we continuously increase the magnetization. The simulations are performed using the relativistic MHD module from the open source, parallel, grid adaptive, MPI-AMRVAC code.
\end{abstract}

\section{Introduction}
Since their discovery in the beginning of the 20th century, astrophysical jets have been extensively observed and studied on multiple scales, ranging from young stellar object jets (YSO jets) and gamma ray bursts (GRBs) to active galactic nuclei jets (AGN jets). Improved instruments allowed for better observations of these peculiar outflows, resulting in the detection of energetic features (e.g. flares), improved resolution and mapping of interesting small-scale regions and information about their overall structure and evolution. The formation, acceleration and collimation of jets was addressed in the beginning analytically and (relatively) recently via numerical simulations, in different regimes, ranging from simple hydrodynamic (HD) to relativistic and general relativistic magnetohydrodynamics (RMHD and GRMHD), with contemporary efforts focusing also on the effect and importance of radiation, e.g. \cite{Sadowski2013}. AGN and GRB jets are relativistic, with Lorentz factors of $\gamma \sim$10 and $\gamma \sim$100 respectively, while YSO jets are non-relativistic, with typical velocities of $\sim 100$ km/s \citep{Woitas2002}.

Most of the scenarios for the large scale jets involve some kind of accretion in a massive central object (i.e. a supermassive black hole) and the presence of a magnetic field. In general, the acceleration of relativistic jets involves the transformation of magnetic energy (Poynting flux) to kinetic, as described with different mechanisms in different scales \citep{BlandfordPayne, BlandfordZnajek}, the efficiency of which is often associated with the shape of the poloidal field lines \citep{Vlahakis2004ApSS, Vlahakis2004ApJ}. The (magnetic, self-) collimation of these outflows is attributed to the hoop stress of the toroidal magnetic field component \citep{Bogovalov1995}, with recent studies focusing on the effect of an external medium, either ambient gas or wind-like outflows from the accretion disk etc., e.g. \citet{Globus2016} and references therein. The interaction with the external medium and causality arguments can possibly explain the stability of jets as mentioned in \cite{Porth2015}.

Analytical work and simulations, in various frameworks and using different physical modules, have been carried out to test the theoretical predictions concerning the launching, acceleration and collimation of jets. For the AGN case we mention e.g. \citet{Casse2004}, \citet{Tchekhovskoy2011}, concerning the launching of jets and the work of \citet{Komissarov2007} where the MHD acceleration of jets was examined. Other studies focused on the development of instabilities in jets, \citep[e.g][]{Baty2002, Bodo2004,Mizuno2012, Bodo2013,Mizuno2014}.

It has also been argued that astrophysical jets are not always homogeneous but rather display a structure in the direction perpendicular to the jet axis. This structure has been observed in terms of velocity (\citet{Mertens2016a, Boccardi2015}) and is believed to exist in different scales, ranging from YSO to AGN jets. Especially for the AGN jet case, the implications on  the radiation emitted from specific sources have been investigated \citep{Giroletti2004, Ghisellini2005}. The above studies resulted in the development of  ``spine \& sheath'' models \citep[e.g][]{Gabuzda2014}, where a jet is believed to consist of a fast inner part (spine) and a slower outer part (sheath).

The effects of this structure have also been considered in some analytical studies, \citep[e.g][]{Bogovalov2001} and via numerical simulations in 2.5D \citep[e.g][]{Meliani2007, Meliani2009}, concerning the axisymmetric stability and in 3D \citep{Singh2016}, concerning the transformation of magnetic to kinetic energy due to kink instability. These papers referred to AGN jets whereas \citet{Matsakos2008, Matsakos2009, Tesileanu2014} examined a two-component YSO jet.

Our primary task in this paper is to examine the effect of this structure to the transverse jet stability and specifically focus on non-axisymmetric instabilities induced by differential rotation, as described in \citet{Meliani2009}. We also note that \citet{Matsumoto2013} argued that in the absence of magnetic field, Rayleigh-Taylor instabilities might still occur in a relativistic jet.

The assumption of a purely poloidal magnetic field will be relaxed, introducing a helical field, while the translational symmetry along the axis of the jet will be maintained for the 2.5D setups. The existence of a toroidal magnetic field component is suggested by \cite{Gabuzda2014,Gabuzda2015}, where different Faraday rotation measures (RM) were examined for a selection of AGN jets. The main part of our 2.5D work will involve a parametric study using different values of magnetization and some extreme scenarios involving a slower inner jet and a slowly rotating jet. The 3D work will only be limited to some extreme cases based on the results obtained by our 2.5D simulations. We assume that the region in question is located sufficiently far from the central source, so the jet is already accelerated and collimated as described in the references above. Therefore, we will only examine cases of kinetically dominated jets. The aim is to see if a helical magnetic field is able to stabilize the jet against instabilities induced by differential rotation.
 
The results of this research may be of use in the FRI/FRII dichotomy \citep{Fanaroff1974}, concerning the morphology of radio jets. Our cases will use standard parameters suitable for AGN jets, but in principle one can examine a similar problem in other astrophysical objects known to host jets (e.g. GRBs). 

\section{Initial state of the jet}
We will follow the same recipe for the overall configuration of the jet and the normalization as described in \citet{Meliani2009}. Velocity is normalized in units of $c$ (and we set $c=1$), distance is measured in $pc$ and mass is normalized to proton mass $m_p$. 

First we fix the (outer) radius of the jet at $R_{out} = 0.1pc$ and the inner radius is arbitrarily chosen to be $R_{in}=R_{out}/3$. These values are consistent with \citet{Biretta2002}, where the opening angle of the M87 jet was found $\sim 10^{\circ}$ at a distance of  $\sim 4 pc$ and the jet is almost fully collimated at the same scale .The next step is to define the initial conditions for the density, velocity, magnetic field and pressure profiles of the jet (inner \& outer regions) and the surrounding, external medium.

We examine a two-component jet in which the two parts of the outflow are differentially rotating, with an initial toroidal velocity profile set as:
\begin{equation} \label{eq: vfprof}
V_{\phi}(R) = \begin{cases}
v_{\phi in} \Bigg(\displaystyle \frac{R}{R_{in}} \Bigg)^{\alpha_{in}/2}, & R\leq R_{in} \\
v_{\phi out} \Bigg(\displaystyle \frac{R}{R_{in}} \Bigg)^{\alpha_{out}/2}, & R_{in} < R < R_{out} \end{cases}
\end{equation}
where we assume $\alpha_{in}=0.5$ for the inner jet and $\alpha_{out}=-2$ for the outer jet. In contrast with \citet{Meliani2007, Meliani2009}, where a profile with a jump at the interface of the two components was used, we assume that in our case the toroidal velocity component is continuous at $R=R_{in}$. The continuity of the toroidal velocity profile is implied by \cite{Mertens2016b}. Initially we will impose this condition by selecting $v_{\phi in}=v_{\phi out}=0.01$ for the main part of our work, but later on we will also examine cases in the non-rotating limit, using $v_{\phi in}= v_{\phi out} = 10^{-6}$.  

\begin{figure*}
\subfloat[Log(Density)]
{\includegraphics[width=0.775\columnwidth, height=3.75cm]{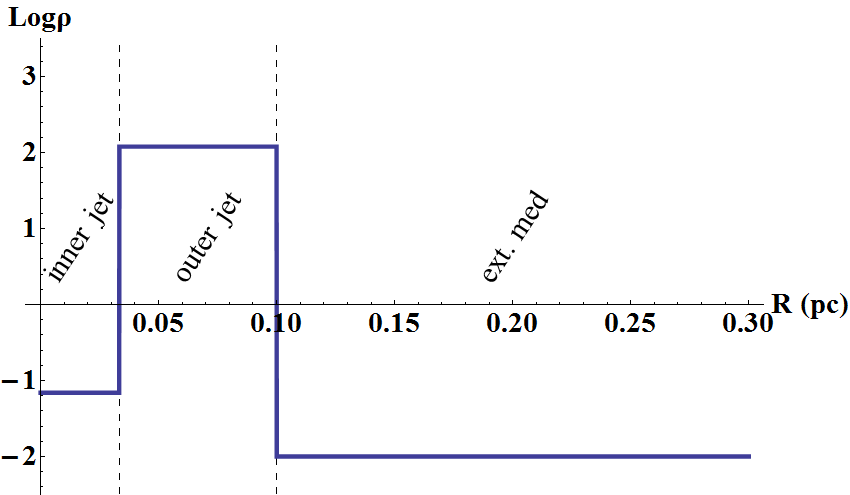} \label{fig: rho}} \hspace{2.5cm}
\subfloat[Toroidal velocity]
{\includegraphics[width=0.775\columnwidth, height=3.75cm]{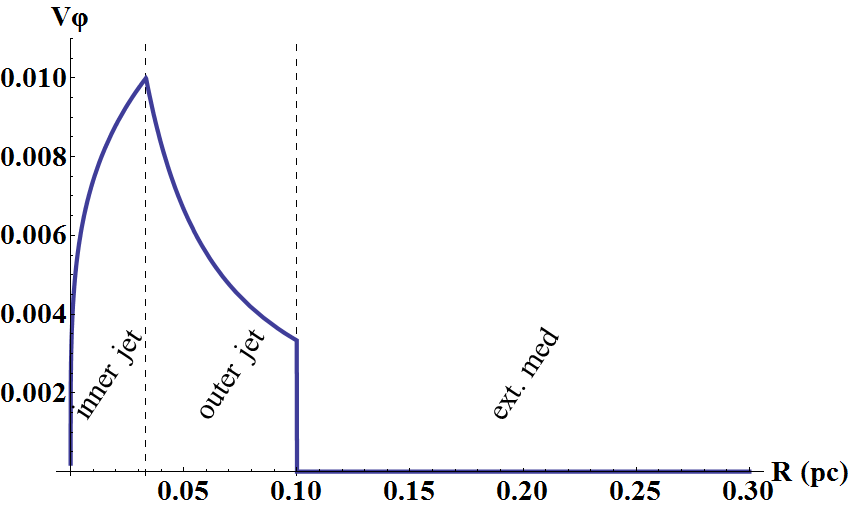}}

\subfloat[Lorentz factor] 
{\includegraphics[width=0.775\columnwidth, height=3.75cm]{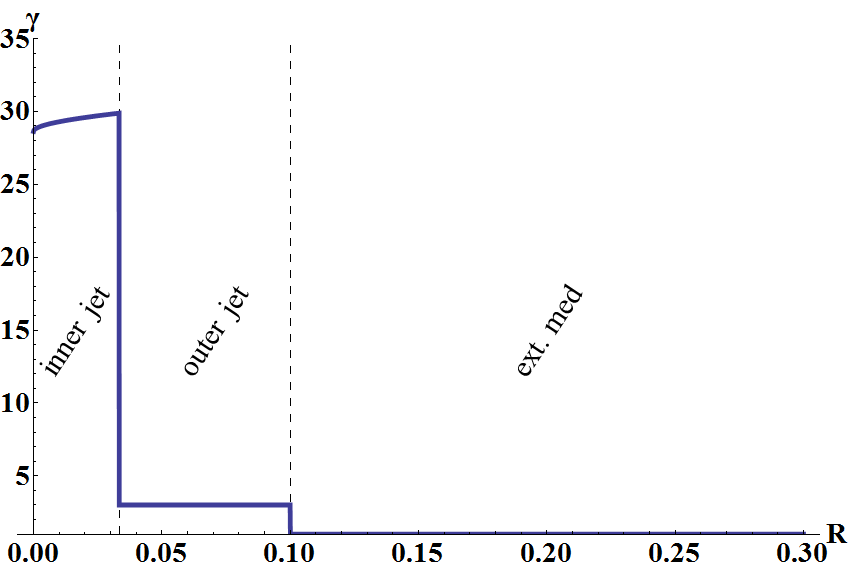} \label{fig: Lorinit}} \hspace{2.5cm}
\subfloat[Magnetic Fieldlines]
{\includegraphics[width=0.775\columnwidth, height=3.75cm]{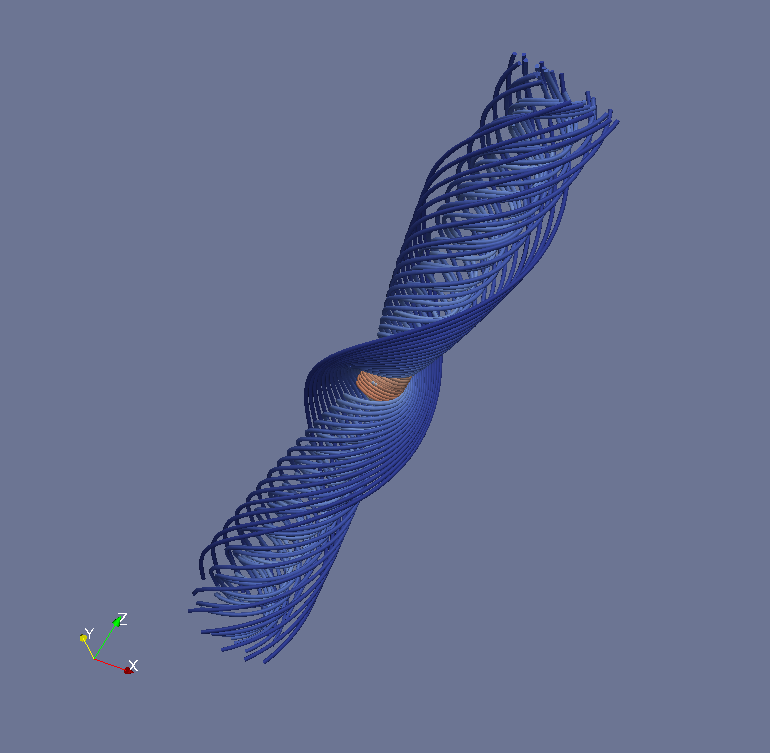} \label{fig: fieldlines}}
\caption{Clockwise from top left: Initial conditions for density, toroidal velocity, magnetic fieldlines and Lorentz factor for the inner \& outer part of the outflow. Concerning the fieldlines, the colour scale represents the magnitude of the magnetic field, so the inner jet is displayed in ``warm'' colours, the outer jet in ``cold'', while the external medium is unmagnetized.}
\end{figure*}

The dominant velocity component of such jet is poloidal and set to a constant value, different for each component. Again the main part of our work will assume $v_z$ with a corresponding Lorentz factor of $\gamma_{z,in} = 30$ for the inner jet and $\gamma_{z,out}=3$ for the outer jet, which are meaningful values for AGN jets \citep{Giroletti2004}. Later on we will decrease the Lorentz factor of the inner jet to $\gamma \simeq 10$. As the toroidal velocity component is significantly smaller than the poloidal, the above values are approximately equal to the total Lorentz factor (see Fig.~\ref{fig: Lorinit}).
  
For the magnetic field, we assume a uniform poloidal component, constant in each part of the jet ($b_{z in}, b_{z out}$) and a toroidal component of the same form as in Eq. \ref{eq: vfprof}:
\begin{equation} \label{eq: bfiprof}
B_{\phi}(R) = \begin{cases}
b_{\phi in} \Bigg(\displaystyle \frac{R}{R_{in}} \Bigg)^{\alpha_{in}/2}, & R\leq R_{in} \\
b_{\phi out} \Bigg(\displaystyle \frac{R}{R_{in}} \Bigg)^{\alpha_{out}/2}, & R_{in} < R < R_{out} \end{cases}
\end{equation}
where the constants $b_{\phi in}, b_{\phi out}$ are defined by fixing the maximum magnetization of the jet, $\sigma= B_{\phi}^2 / (\gamma^2 \rho)$ at $R=R_{in}$. Similarly to the definition of the toroidal velocity, we select $b_{\phi in} = b_{\phi out}$, while the poloidal magnetic field will still be discontinuous at $R=R_{in}$. The initial 3D reconstruction of the fieldlines of this helical magnetic field are shown in Fig.~\ref{fig: fieldlines}. For completeness, we include the profiles of $V_z$ and $B_z$ with $R$:
\begin{equation} \label{eq: vzprof}
V_{z}(R) = \begin{cases}
v_{zin}, & R\leq R_{in} \\
v_{zout}, & R_{in} < R < R_{out} \end{cases}
\end{equation}   
\begin{equation} \label{eq: bzprof}
B_{z}(R) = \begin{cases}
b_{zin}, & R\leq R_{in} \\
b_{zout}, & R_{in} < R < R_{out} \end{cases}
\end{equation} 
where $v_{zin},v_{zout},b_{zin},b_{zout}$ are constants.

The jet is surrounded by a static external medium, with a density roughly of the same order of magnitude as of the inner jet.We set the number density of the surrounding medium to $n_{med}=10^{-2} cm^{-3}$. The jet density values can be calculated using constraints from observations, namely the kinetic luminosity flux for a typical radio loud galaxy ($\sim10^{46}$ ergs/s) and the ratio of the flux carried by each jet component. If the inner part carries only 1 \% of the flux, then the density values for the inner and outer jet are $\rho_{in}= 6.92 \rho_{med}$ and $\rho_{out} = 119.94 \ 10^2 \ \rho_{med}$, resulting in an initial density ratio of $\sim 10^4$ between the components. In all cases, the jets are over-dense. This is related to the fact that we are modelling the jet at a sub-parsec region, where the external medium could result from a disc wind and/or a hot medium in the inner region of the AGN. The chosen value of density for the surrounding medium is also reasonable for a jet cocoon \citep{Meliani2008}.

The density is constant in each part of the jet (Fig.~\ref{fig: rho} and equation~(\ref{eq: densprof})).
\begin{equation} \label{eq: densprof}
\rho(R) = \begin{cases}
\rho_{in}, & R\leq R_{in} \\
\rho_{out}, & R_{in} < R < R_{out} \\
\rho_{med}, & R >R_{out}
 \end{cases}
\end{equation} 
The relativistic equivalent of Rayleigh's criterion for rotational stability is that the angular momentum flux must increase with the radial distance $R$. The angular momentum flux is given by the formula:
\begin{equation}
I \ = \ \gamma \displaystyle \frac{\rho + \displaystyle \frac{\Gamma}{\Gamma-1}}p{\rho}V_{\phi}R- \displaystyle \frac{B_p}{\gamma \rho V_p}R B_{\phi}
\end{equation}
and thus if $B_{\phi}=0$ and $V_{\phi}$ as in equation~(\ref{eq: vfprof}), the inner jet is stable (as $\frac{dI}{dR} > 0$) and the outer jet is marginally stable, as $\frac{dI}{dR}=0$. The interface between the two components is still unstable, as the angular momentum changes significantly at $R=R_{in}$. In our study, all $B_z$ values still satisfy the criterion.
 
We assume in addition total pressure equilibrium between the two interfaces (inner \& outer jet, outer jet \& external medium) at $t=0$. The steady state momentum equation is \citep[e.g.][]{Heyvaerts2003}:
\begin{equation}
\gamma \rho (\vec{v} \cdot \nabla) (\xi \gamma \vec{v}) = - \nabla P + \vec{J} \times \vec{B}+\rho_e \vec{E}
\end{equation}
which is equivalent to equation 23 in \cite{Mobarry1986}, ignoring the geometric/gravity terms. Assuming a polytropic equation of state, the differential equation describing the variation of the total pressure $P_{tot}$ along the radial direction is:
\begin{gather} \label{eq: totPeq}
\begin{split}
& \frac{d P_{tot}}{d R} - \frac{\Gamma}{\Gamma-1} \frac{\gamma^2 V_{\phi}^2}{R}P_{tot} = \frac{\gamma^2 V_{\phi}^2}{R} \Bigg(\rho-\frac{\Gamma}{\Gamma-1}\frac{B_z^2}{2}\Bigg) \\
&+ \frac{1}{R}\Bigg[ 1+\frac{\Gamma \gamma^2 V_{\phi}^2}{2(\Gamma-1)} \Bigg] \Bigg[ -B_{\phi}^2+ \Bigg( B_{\phi}V_z-V_{\phi}Bz \Bigg)^2 \Bigg]
\end{split}
\end{gather} 
where $P_{tot} = p + \displaystyle \frac{B^2-E^2}{2}$ is the total pressure. Using equations  ~(\ref{eq: vfprof}), ~(\ref{eq: bfiprof}) and ~(\ref{eq: totPeq}), we find that the thermal pressure $p$ varies with $R$ as: 
\begin{gather} \label{eq: preseq}
\begin{split}
&p = \zeta \Bigg[ 1 - \tilde{a} \Bigg( \frac{R}{R_{in}} \Bigg)^{\alpha} \Bigg ]^{-\displaystyle \frac{\Gamma}{\alpha(\Gamma-1)}} - \frac{\Gamma-1}{\Gamma}\rho \\ 
&- \frac{(\Gamma-1)(\alpha+2)}{2 \tilde{a} [\alpha(\Gamma-1)+\Gamma]} \Bigg[-b_{\phi}^2+ \Bigg( b_{\phi}V_z-v_{\phi}B_z \Bigg)^2 \Bigg]  \Bigg(1-\tilde{a} \Bigg(\frac{R}{R_{in}}\Bigg)^\alpha \Bigg)
\end{split}
\end{gather}
where $\tilde{a} = \frac{v_{\phi}^2}{1-v_z^2}$ and the parameter $\alpha$ is $\alpha_{in}$ and $\alpha_{out}$ for the inner and outer jet respectively. The integration constant $\zeta$ ($\zeta_{in}, \zeta_{out}$, depending on the component) is obtained from the boundary conditions (the thermal pressure on the axis, $p_o$) and the total pressure matching at the interface of the two components. The full expressions for these constants can be found in the appendix. The effective polytropic index $\Gamma$ is initially approximately 4/3 for the inner jet and the external medium and 5/3 for the outer jet, which correspond to relativistically hot and cold outflows respectively. 

Following \citet{Meliani2009}, we will later use a Synge equation of state for the simulations, since mixing of matter from both parts of the jet is present. This does not contradict the earlier polytropic approximation since $\Gamma_{eff,in}$ and $\Gamma_{eff,out}$ are constant in the initial state.
\section{Numerical setup \& different cases}\label{sec: setups}
We will examine different configurations of two component jets, surrounded by a static medium,  including a toroidal magnetic field component in both parts of the jet. This results in non-zero magnetization for both components, with each sub-case using a different value for the maximum magnetization. All the chosen values of magnetization, although underestimated, correspond to a kinetically dominated jet, as $\sigma \leq 0.1$, which is in agreement with the assumption that far from the central engine, AGN jets have already been accelerated and collimated. 

We select values of magnetization ranging from $\sigma = 0.001$ to $\sigma = 0.1$. Every 2.5D run is performed on a cartesian  domain with dimensions $-0.3pc < x,y < 0.3pc$. We use the relativistic MHD module of the MPI-AMRVAC code \citep{Keppens2012, Porth2014} selecting an HLLC solver and a third order limiter \citep{Cada2009}. The base resolution of each simulation is $128 \times 128$ with 3 levels of adaptive mesh refinement (AMR) used, achieving an effective resolution of $512 \times 512$ (thus covering $\sim 0.001pc$ per cell in the finest resolution). Higher resolution runs were also performed, where fine structure effects naturally appeared, with the overall image remaining the same. Every simulation covers a time range of t=60 or 195.8 years, which is roughly equal to 3 rotations of the inner jet. The boundary conditions are open in every side of the computational box, with restrictions clipping any inflow. We also use two different tracers, one for each jet component, to detect any mixing between the two parts of the outflow. 

The 2.5D simulations can be categorized as follows: (I) simulations for different values of $\sigma$ where we chose for the inner jet $\gamma = 30$ (II) Cases using the lowest and highest values of magnetization values of (I) with $\gamma = 10$ for the inner jet and (III) Cases using the lowest and highest values of magnetization assuming a very slow rotation for the jet. In all our 2.5D cases, the thermal pressure on the axis is set to $p_o=2$. When the initial Lorentz factor of the inner jet is reduced to $\gamma = 10$, the total kinetic luminosity flux is reduced by less than $\sim$1\%. Using the same configuration for the outer jet, the density ratio between the components remains constant and thus the main contribution to the kinetic flux is still due to the outer jet, in both cases.

For the 3D simulations, we chose periodic boundary conditions on the $z$ axis, while the computational domain is a cartesian box with dimensions $-0.3pc < x,y < 0.3pc$, $0 < z < 1pc$ and a resolution of $256^3$, without AMR. We examine the most interesting cases of the 2.5D scenarios, as explained in the relevant section. The thermal pressure on the axis is set to $p_o=2.3$. For simplicity we consider no scaling with $z$ for all physical quantities. Although most studies assume $B_{\phi} \sim 1/z$ and $B_z \sim 1/z^2$, we expect little difference in the small scale examined here. Our 3D simulations therefore correspond to a cylindrical jet.

In some cases, we notice initially the formation of four ``arms'', which is due to the fact that we use a Cartesian grid and thus modes with $m=4$ are being favoured. This can be suppressed by exciting specific modes in the perturbations of the radial velocity. Two different types of perturbations are used, either with $\nabla \cdot \vec{v}$ or perturbations with selected $m$ \citep{Rossi2008}. The kind of perturbation will be mentioned in each different case. 
\begin{figure}
\centering
\includegraphics[width=0.99\columnwidth, height=5.5cm]{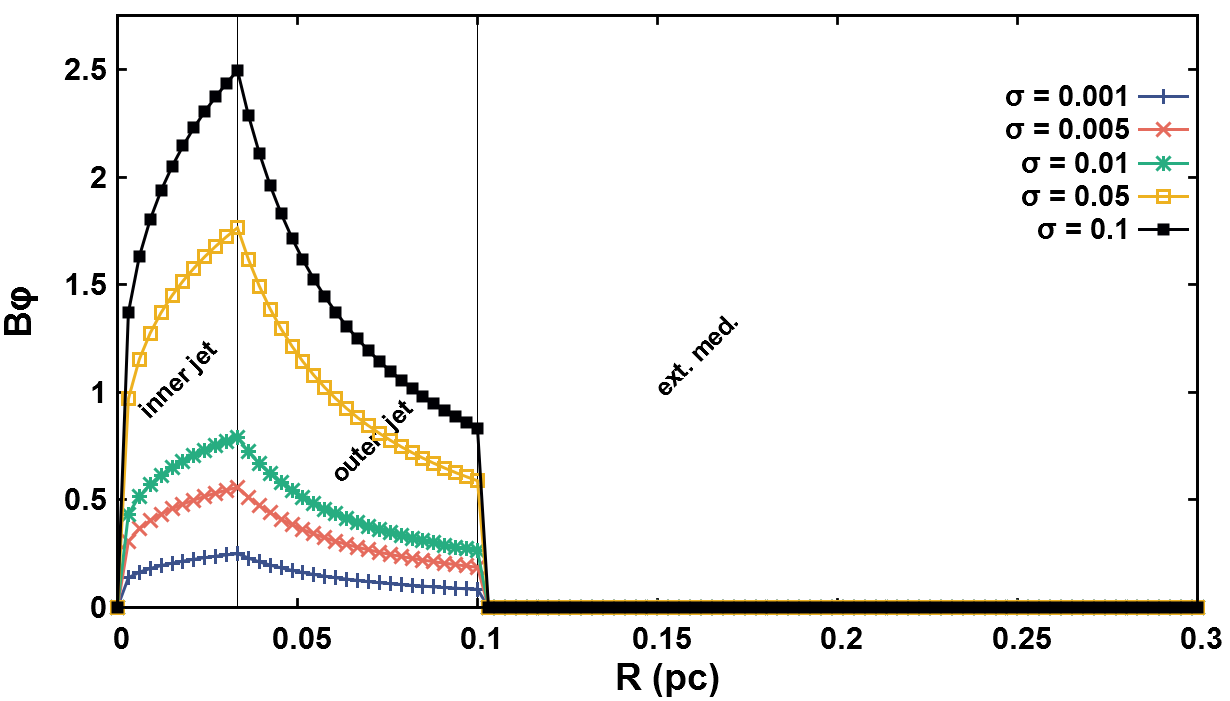} \\
\caption{Toroidal magnetic field magnitude for five different values of magnetization $\sigma$, also mentioned in Table \ref{tab: table_cases} \label{fig: bfcases}}
\end{figure}
\begin{table*}
\centering
\caption{Key parameters for the examined cases. The average Lorentz factor for the inner jet and the effective radius of the entire jet after 3 rotations are given in the last two columns.}
\label{tab: table_cases}
\begin{tabular}{cccccccc} 
\hline
Case \#  & $\sigma_{max}$ & $\gamma_{in,0}$ & $v_{\phi}$ &$B_{z,in}$ & $B_{z,out}$  & $\gamma_{in,avg}$  & $R_{eff, jet}$ ($pc$)\\
\hline  \hline
1        &  0.001   &   30   &   0.01             &   1.228   &   1.039    &     10.9    &  0.144 \\
2        &  0.005   &   30   &   0.01             &   0.549   &   1.039    &     13.5    &  0.134 \\
3        &  0.01     &   30   &   0.01             &   0.388   &   1.039    &     15.4    &  0.129 \\
4        &  0.05     &   30   &   0.01             &   0.174   &   1.039    &     21.4    &  0.116 \\
5        &  0.1       &   30   &   0.01             &   0.123   &   1.039    &     22.2    &  0.116 \\
\hline
6        &  0.001   &   10   &   0.01             &   0.429   &   1.039    &       7.5    &  0.125\\
7        &  0.1       &   10   &   0.01             &   0.043   &   1.039    &       9.8    &  0.12\\
\hline
8        &  0.001   &   10   &   $10^{-6}$   &  0.429   &   1.039    &      27.0    &  0.132\\
9        &  0.1       &   10   &   $10^{-6}$   &  0.043   &   1.039    &      26.7    &  0.110\\
\hline
\end{tabular}
\end{table*}
\section{Increasing the magnetization} \label{sec: results1}
We first present the results for the cases with a Lorentz factor of $\gamma = 30$ for the inner jet, where we follow the evolution of the density and the Lorentz factor with time. A wide range of different magnetization values is used and the key values are summarized in Table~\ref{tab: table_cases}. We will present in more detail three representative cases, with $\sigma = 0.001$, $0.01$ and $0.1$ while the rest will be used to examine the existence of trends in the evolution of the average Lorentz factor and the effective radius of the jet with $\sigma$. A comparison of the strength of the toroidal magnetic field, in normalized units for our work, for different values of $\sigma$ is shown in Fig~\ref{fig: bfcases}. The initial state of the jet in terms of density is the same for all simulations and thus it will be presented only once in the first case.
\subsection{Case 1: $\sigma=0.001$ , $\gamma_{in} \simeq 30$} \label{sec: sig0_001}
From a physical point of view, we expect this case to be the most unstable one, since the hoop stress is minimum and the small magnetization value makes it effectively a (relativistic) hydrodynamic setup, somewhat similar to Case D studied in \citet{Meliani2009} (since $B_z$ only contributes to the total pressure). In this case we follow equation 3 of \citet{Rossi2008} in the choice of the initial radial velocity perturbation, choosing $m=3$ while $\omega_l,b_l,l$=0. 

\begin{figure}
\centering
\includegraphics[trim={0.5cm 0.7cm 0.5cm 0.5cm}, clip, scale=0.175]{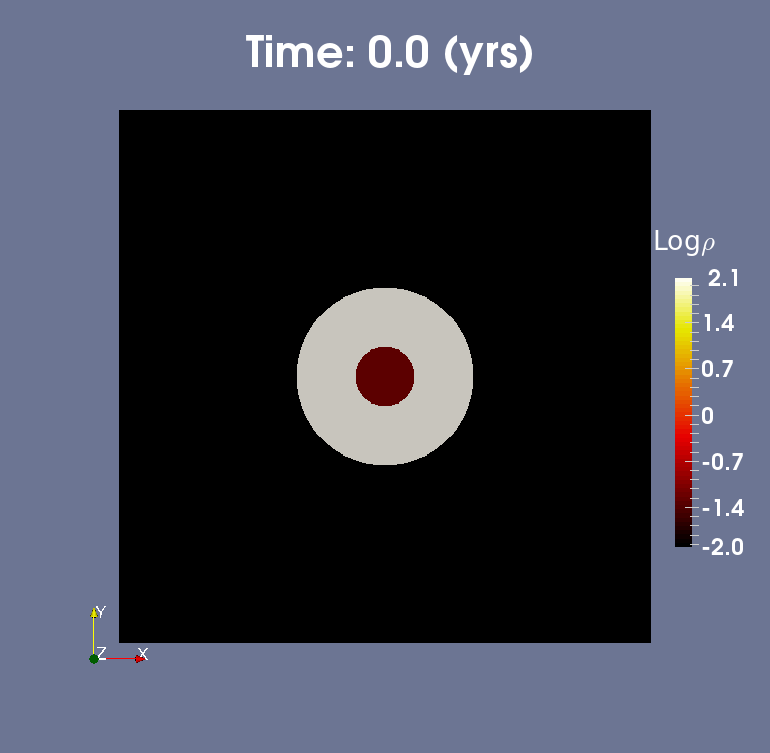} \\
\includegraphics[trim={0.5cm 0.7cm 0.5cm 0.5cm}, clip, scale=0.175]{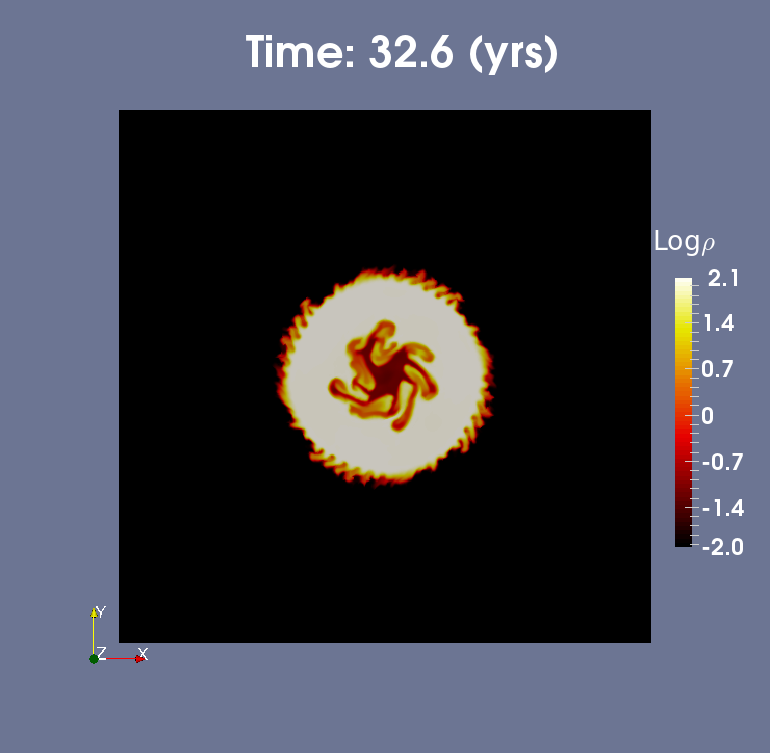} \\
\includegraphics[trim={0.5cm 0.7cm 0.5cm 0.5cm}, clip, scale=0.175]{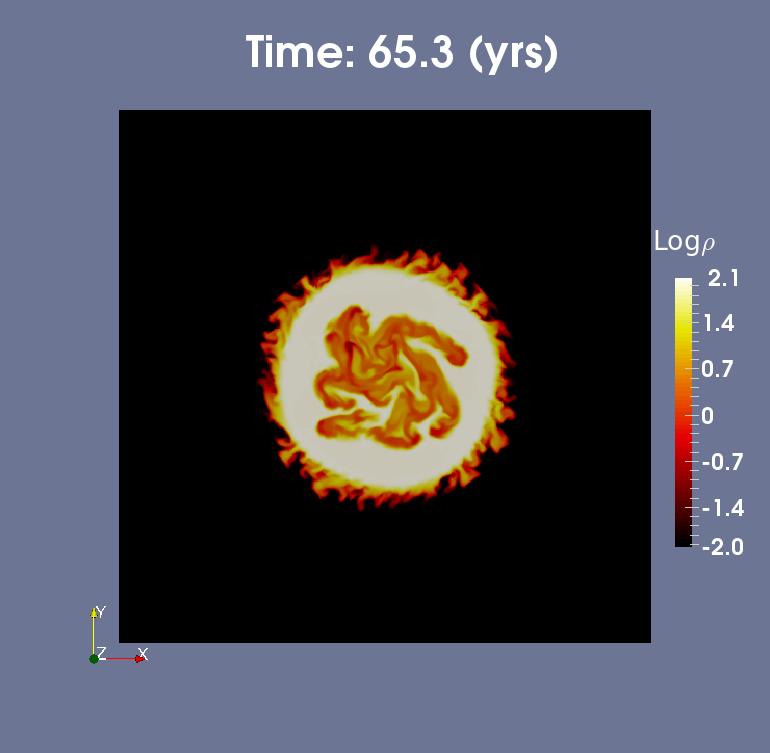}  \\
\includegraphics[trim={0.5cm 0.7cm 0.5cm 0.5cm}, clip, scale=0.175]{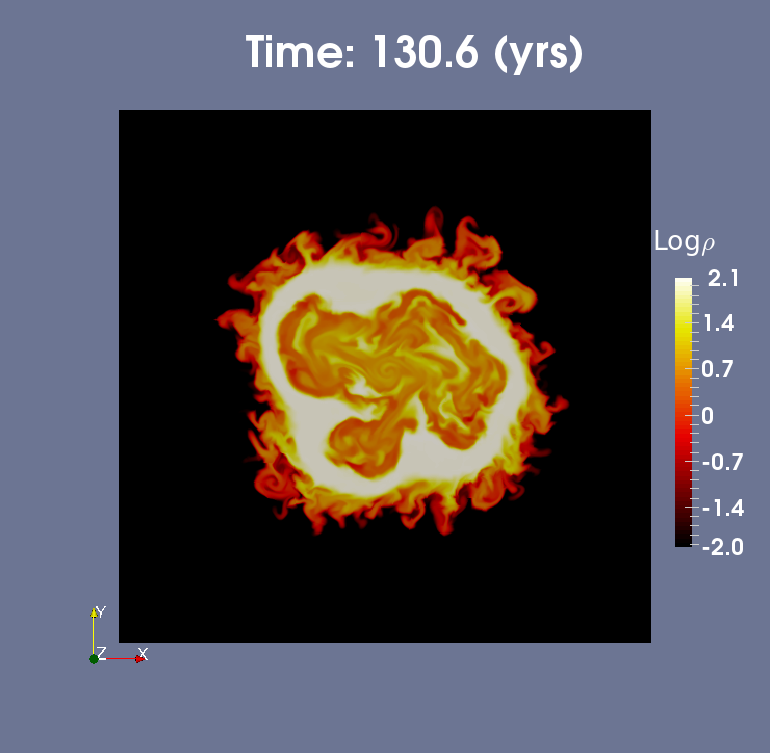}  \\
\includegraphics[trim={0.5cm 0.7cm 0.5cm 0.5cm}, clip, scale=0.175]{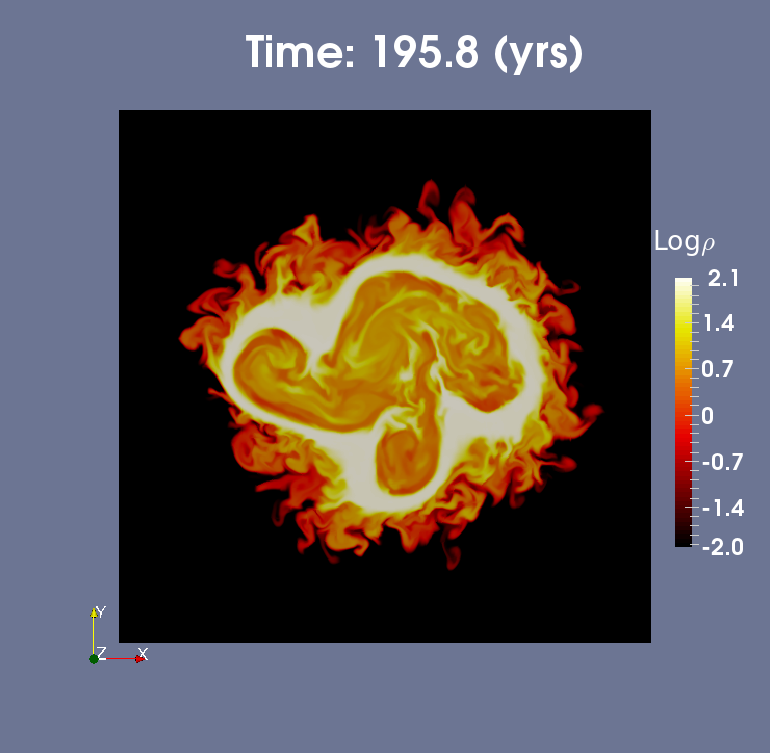}  \\
\caption{Proper density (log) for Case 1 (maximum $\sigma=0.001$, $\gamma_{in} \simeq 30$). Snapshots after 0, 0.5, 1, 2 and 3 rotations of the inner jet. \label{fig: sig0_001dens}} 
\end{figure}

Initially, the centrifugal force dominates over the magnetic tension, a fact which is reflected in the quick expansion of the surface of the jet (both for the inner and the outer part), shown in Fig.~\ref{fig: sig0_001dens}. Even though the toroidal velocity transition is continuous at $R_{in}$ --and not a jump, as in the original study of \citet{Meliani2009}, some Kelvin-Helmholtz instabilities develop in the early stages of the simulation, but are quickly overcome by a Rayleigh-Taylor type instability.
\begin{figure}
\centering
\subfloat[Lorentz factor, t=0]
{\includegraphics[trim={0.5cm 0.7cm 0.5cm 0.5cm}, clip, scale=0.155]{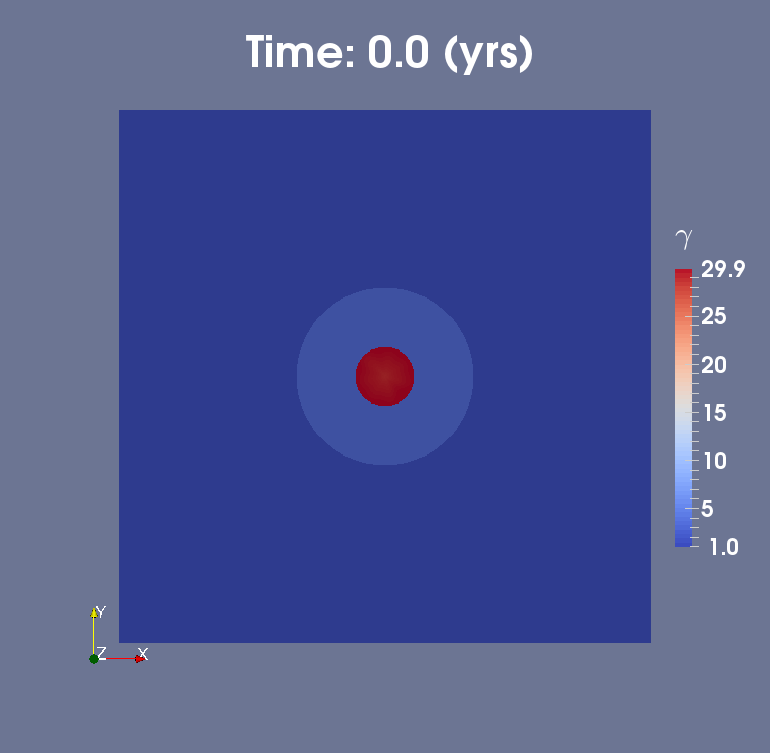}}
\subfloat[Lorentz factor, t=3 rotations]
{\includegraphics[trim={0.5cm 0.7cm 0.5cm 0.5cm}, clip, scale=0.155]{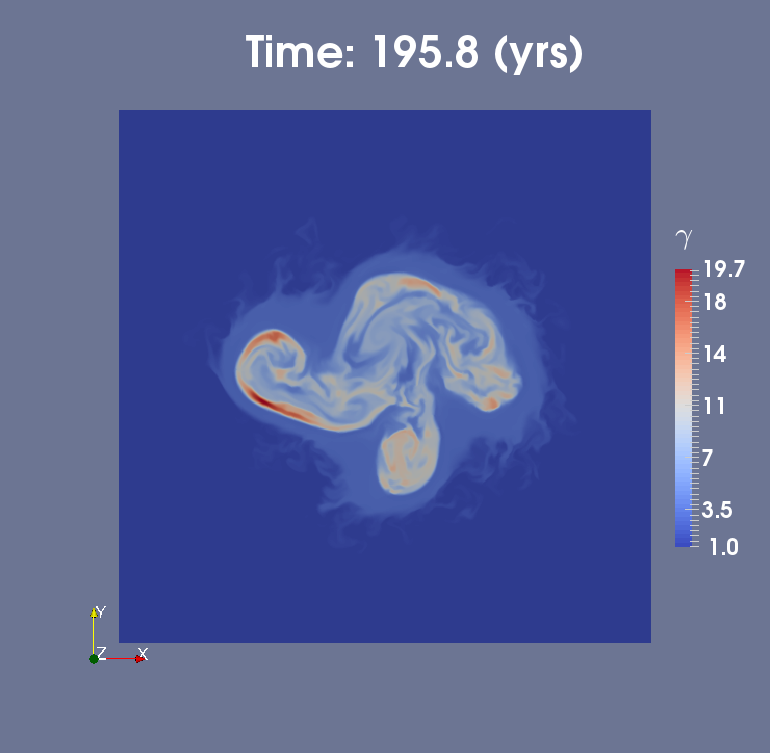}}
\caption{Lorentz factor for Case 1 (maximum $\sigma=0.001$, $\gamma_{in} \simeq 30$) after 3 rotations of the inner jet. \label{fig: sig0_001lfac}} 
\end{figure}
In the inner part, we notice the formation of ``arms'' quite early (around half rotation time of the inner jet), a feature which persists throughout the evolution of the jet. Apart from the inner jet material getting mixed in the outer jet by Rayleigh-Taylor formation, there is also inflow from the outer component into the inner jet region. This inflow is of high density and low angular momentum, which contributes to the deceleration of the jet. The mixing starts after $\sim 0.25$ rotations of the inner jet, which becomes clear after $\sim 1$ rotation. In the end, a mixed region is formed in the central part, while the high-density outer part of the jet is pushed outwards and the jet as a whole expands. The density ratio between the final state and the initial value for the inner jet can reach the order of $\sim 100$.

The dominant instability in this simulation is thus a relativistically enhanced Rayleigh-Taylor type, as discussed in \citet{Meliani2009}, with Kelvin-Helmholtz effects remaining present mostly early in the simulation and restricted in the boundary between the two jet components. The magnetic tension remains weak throughout the evolution and is not sufficient to constrain the jet expansion.

The distribution of Lorentz factor in the final state is given in Fig.~\ref{fig: sig0_001lfac}. The jet clearly de-collimates after one rotation time, finally reaching an effective radius of $R_{eff}\simeq 0.144 pc$ and the average Lorentz factor for the inner jet is decreased to $\sim10.9$.  
\subsection{Case 3: $\sigma=0.01$ , $\gamma_{in} \simeq 30$}
We increase the maximum magnetization by a factor of 10 with respect to the first case, to $\sigma=0.01$. In this case we do not excite specific modes but rather use a zero-divergence perturbation in the radial velocity.
\begin{figure}
\centering
\includegraphics[trim={0.5cm 0.7cm 0.5cm 0.5cm}, clip, scale=0.175]{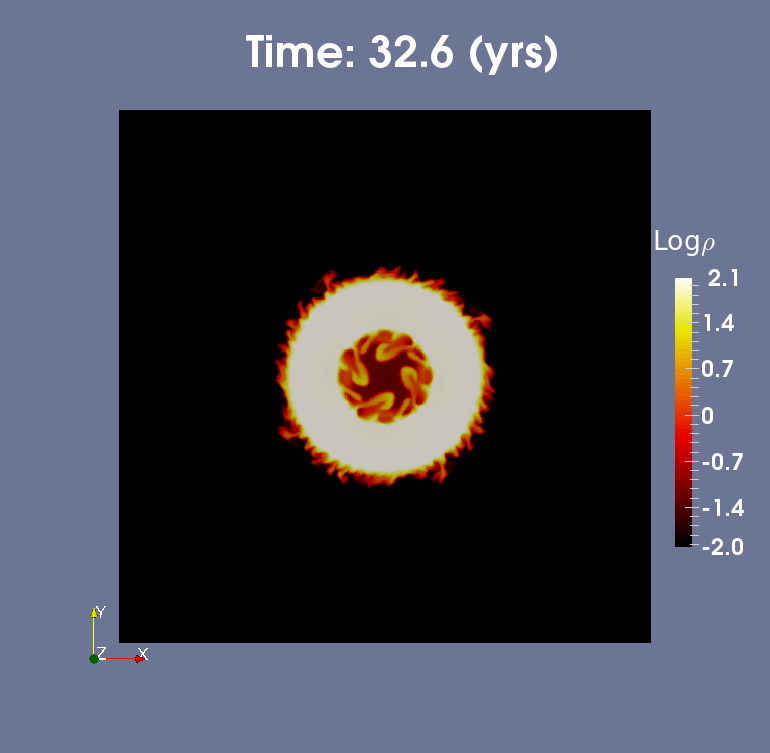} \\
\includegraphics[trim={0.5cm 0.7cm 0.5cm 0.5cm}, clip, scale=0.175]{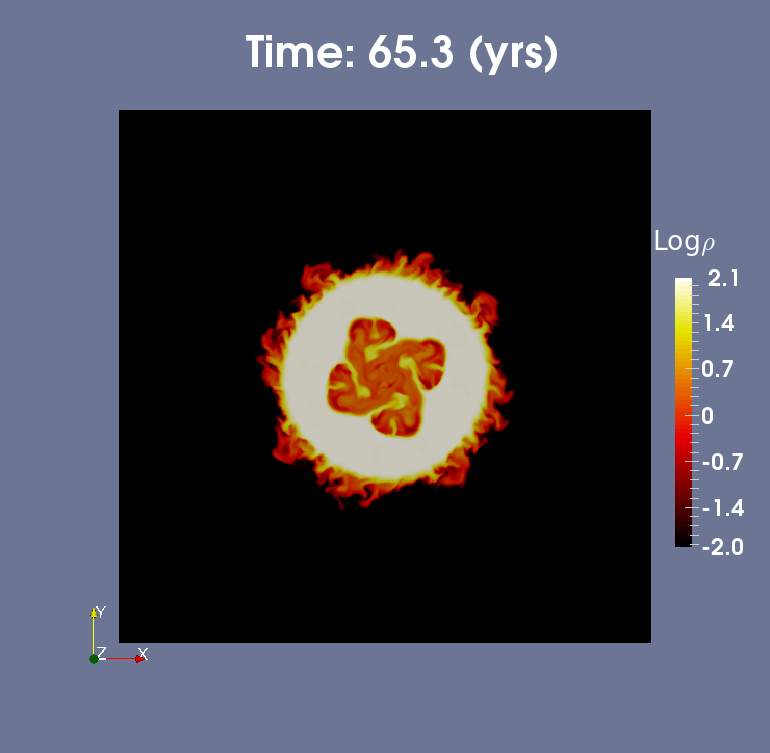}  \\
\includegraphics[trim={0.5cm 0.7cm 0.5cm 0.5cm}, clip, scale=0.175]{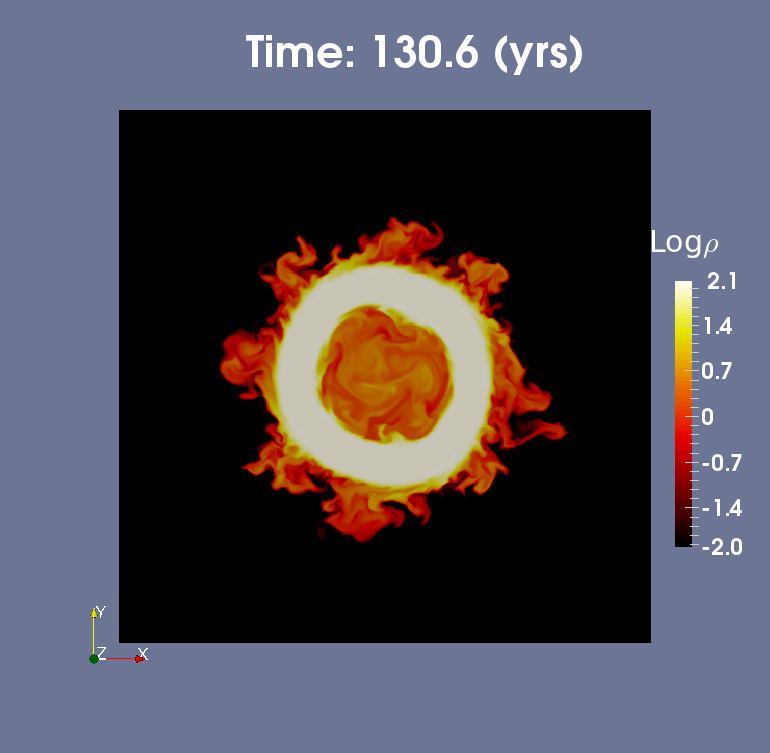}  \\
\includegraphics[trim={0.5cm 0.7cm 0.5cm 0.5cm}, clip, scale=0.175]{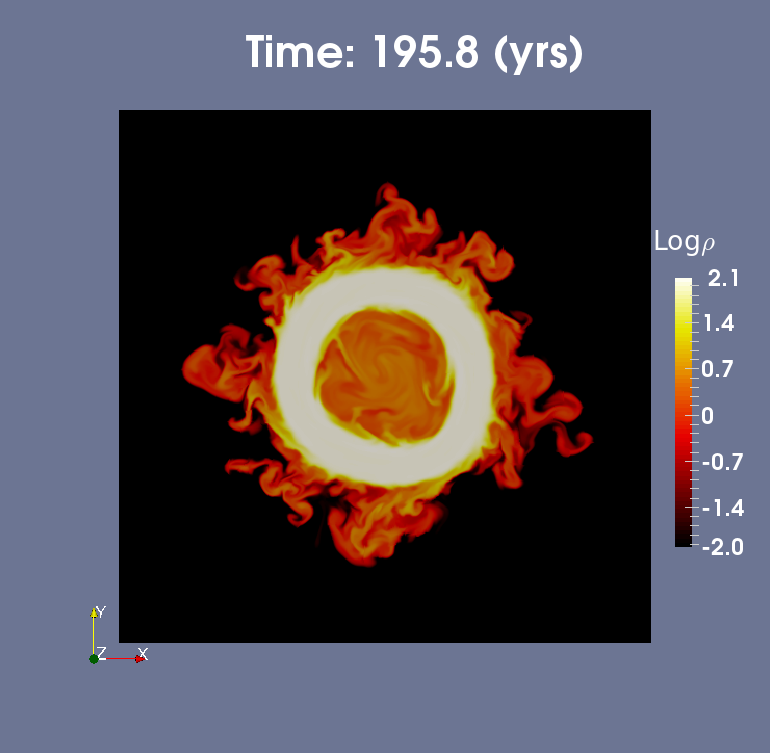} \\
\includegraphics[height=4.75cm,width=0.92\columnwidth]{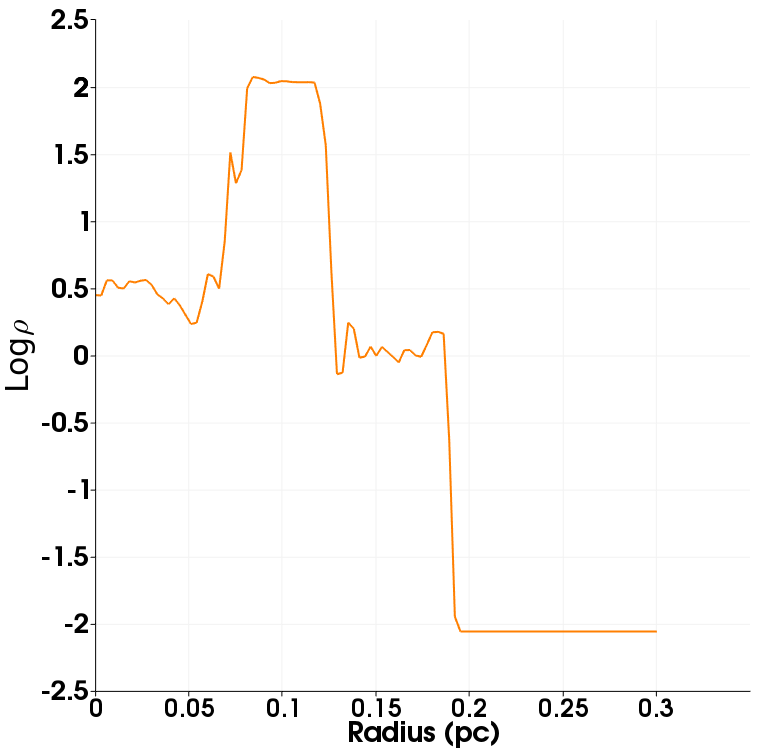} 
\caption{Proper density (log) for Case 3 (maximum $\sigma=0.01$,  $\gamma_{in} \simeq 30$). Snapshots after 0.5, 1, 2 and 3 rotations of the inner jet. The bottom image shows the distribution along the $x$ axis at $t=3$ (i.e. 195.8 yrs). \label{fig: sig0_01dens}} 
\end{figure}
\begin{figure}
\centering
{\includegraphics[trim={0.5cm 0.7cm 0.5cm 0.5cm}, clip, scale=0.155]{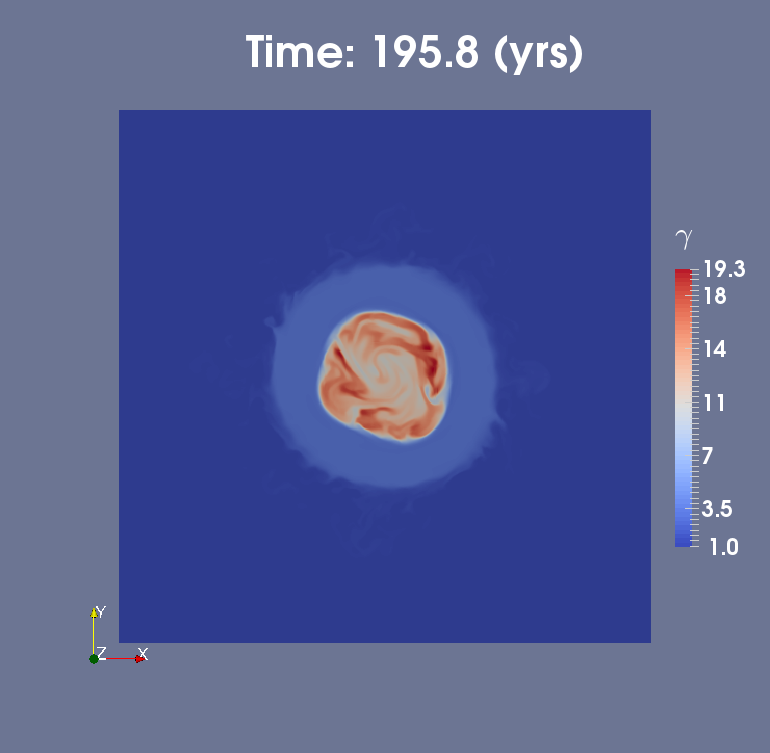}}
\caption{Lorentz factor for Case 3 (maximum $\sigma=0.01$,  $\gamma_{in} \simeq 30$) after 3 rotations of the inner jet. \label{fig: sig0_01lfac}} 
\end{figure}
The first stages of the evolution resemble the previous case, at least up to $\sim 0.5$ rotations, with low density material from the inner jet moving outwards, due to the centrifugal force and dense, low angular momentum material falling inwards. We now notice the numerical selection of 4 arms effect, favoured by our cartesian grid, seen up to $\sim 1$ rotation time of the inner jet. Later, after $\sim 2$ rotations of the inner jet, the central part of the jet consists of a well mixed region.  After the formation of the mixed region, the ``high'' and ``low'' density components of the outflow can be distinguished once more, with different values compared to the initial state. Although what remains of the outer component does not experience any significant change in density, the inner jet remains in a mixed state, with a density of $\sim 50$ times greater than the initial inner jet density. The evolution of the density with time can be seen in Fig.~\ref{fig: sig0_01dens}. We also present the distribution with the radial distance, e.g. along the x axis. As seen in the previous case, the problem is not axisymmetric, but we can still obtain qualitative information about the mixing.

The increased hoop stress prevents in a sense the development of evident Rayleigh-Taylor type instabilities in large scale, whereas we still notice the development of Rayleigh-Taylor type instability between the outer jet and the environment. As before, any Kelvin-Helmholtz phenomena between the two jet components are quickly suppressed.

The Lorentz factor distribution after 3 rotation times is shown in Fig.~\ref{fig: sig0_01lfac}. Observing the evolution of the Lorentz factor, we notice that even though the effective radius of the inner jet increases, an inner, fast and an outer, slow region can be clearly distinguished in the outflow for the entire evolution shown. The average Lorentz factor after 3 rotations is $13.5$ and the effective radius of the jet $R_{eff}\simeq 0.134pc$.
\subsection{Case 5: $\sigma=0.1$ , $\gamma_{in} \simeq 30$} \label{sec: sig0_1}
Here we examine the case with the maximum value of magnetization in our study, $\sigma = 0.1$. Following the evolution of the previous cases, we expect this configuration to be the most stable one, due to the increased magnetic tension. We use the same kind of perturbation as in the case with $\sigma=0.01$. 

As seen earlier in this work, in the early stages we notice the formation of  ``arms'' and some Kelvin-Helmholtz effects between the two jet components. In this case however, the mixing starts after $\sim 0.5$ rotation and is restricted near the interface of the two jet components. The ``arms'' persist up to $\sim 0.75$ rotation of the inner jet and later form a mixed region in the edge of the inner jet.
\begin{figure}
\centering
\includegraphics[trim={0.5cm 0.7cm 0.5cm 0.5cm}, clip, scale=0.175]{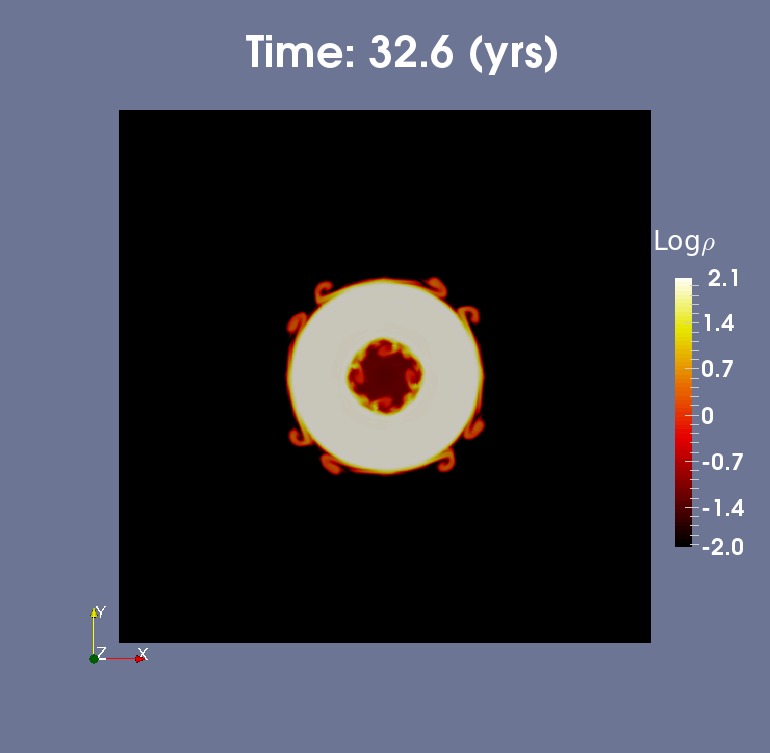} \\
\includegraphics[trim={0.5cm 0.7cm 0.5cm 0.5cm}, clip, scale=0.175]{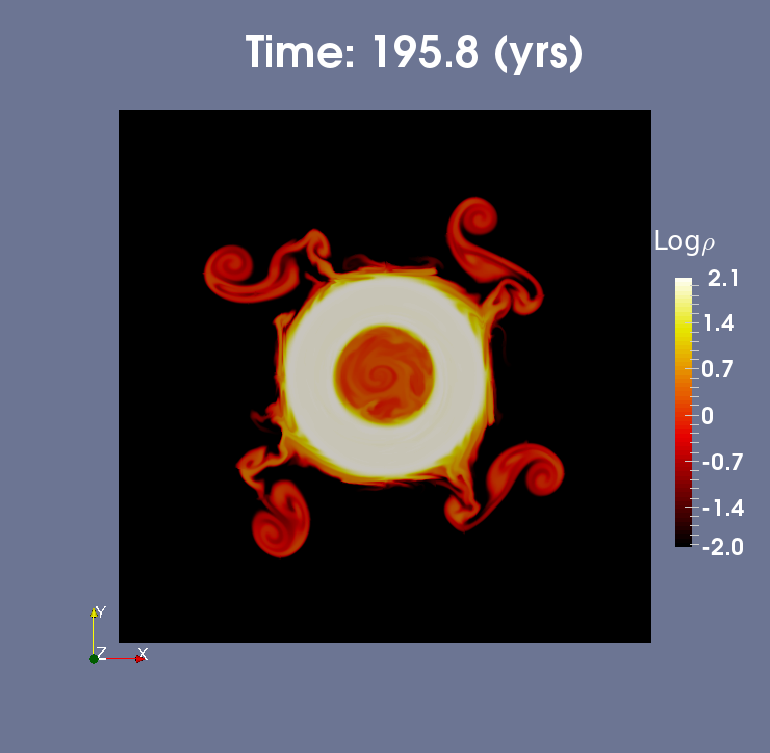} \\
\includegraphics[height=4.75cm,width=0.92\columnwidth]{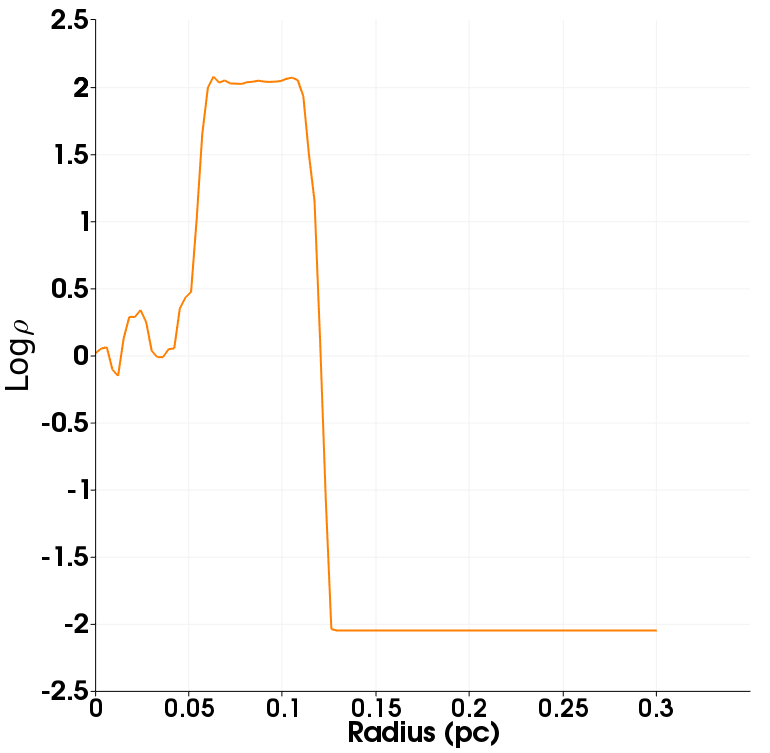} 
\caption{Proper density (log) for Case 5 (maximum $\sigma=0.1$,  $\gamma_{in} \simeq 30$). Snapshots after 0.5 and 3 rotations of the inner jet. The bottom image shows the distribution along the $x$ axis at $t=3$ (i.e. 195.8 yrs). \label{fig: sig0_1dens}} 
\end{figure}
\begin{figure}
\centering
{\includegraphics[trim={0.5cm 0.7cm 0.5cm 0.5cm}, clip, scale=0.155]{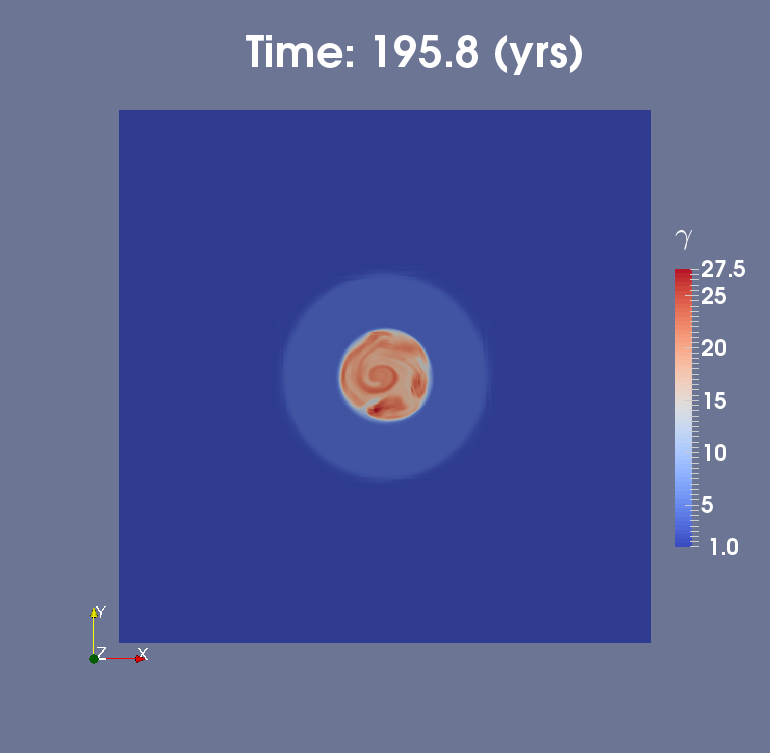}}
\caption{Lorentz factor for Case 5 (maximum $\sigma=0.1$,  $\gamma_{in} \simeq 30$) after 3 rotations of the inner jet. \label{fig: sig0_1lfac}} 
\end{figure}
The mixing is now present in a quite smaller region compared to the previous cases, with the final state resulting in a ``fast jet'' area smaller than the one found in the unstable cases.  Any enhanced Rayleigh-Taylor type phenomenon is quickly (before $\sim 1.5$ rotations of the inner jet) suppressed. 

The final state is in a sense similar to the previous case due the strong magnetic tension, resulting again in a fast, inner and a slow, outer component. We note that even in this case, the inner jet has a higher density compared to the initial state due to the mixing process, with a numerical value $\sim 20$ times greater than the initial value for the inner jet. The distribution of the density over time and the distribution along the $x$ axis can be found in Fig.~\ref{fig: sig0_1dens}. The Lorentz factor after 3 rotations is given in Fig.~\ref{fig: sig0_1lfac}.

The outer jet, in terms of density, remains unaffected and after 3 rotations is still more or less uniform. The effective radius of the inner jet is also in this case increased but the two components are more structured and distinguishable. The average Lorentz factor of the inner jet after 3 rotations is $\sim 22.2$ and the effective radius of the whole outflow is $R_{eff}\simeq 0.116pc$.
\subsection{Average Lorentz factor, Effective radius and mixing}
\begin{figure*}
\centering
\subfloat[$\bar{\gamma}_{in}(t)$]
{\includegraphics[width=0.94\columnwidth, height=4.95cm]{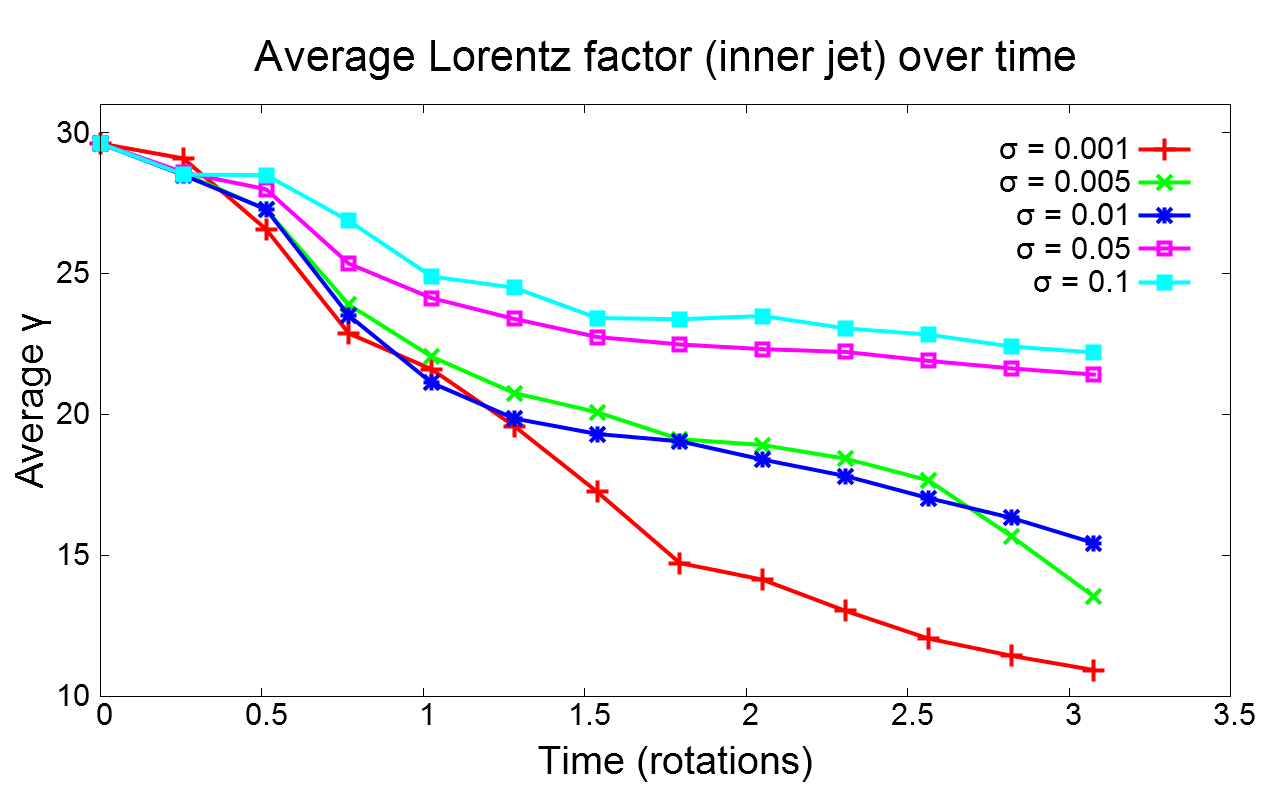}} \hspace{1.5cm}
\subfloat[$\Gamma_{eff}$, $t=0$, all cases]
{\includegraphics[width=0.94\columnwidth, height=4.95cm]{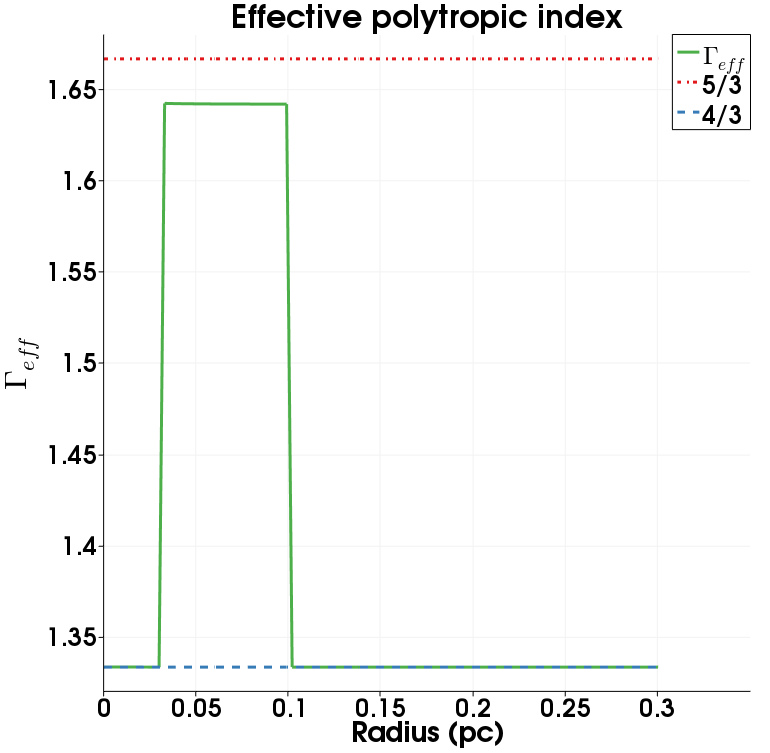}}

\subfloat[$R_{eff in}(t)$]
{\includegraphics[width=0.94\columnwidth, height=4.95cm]{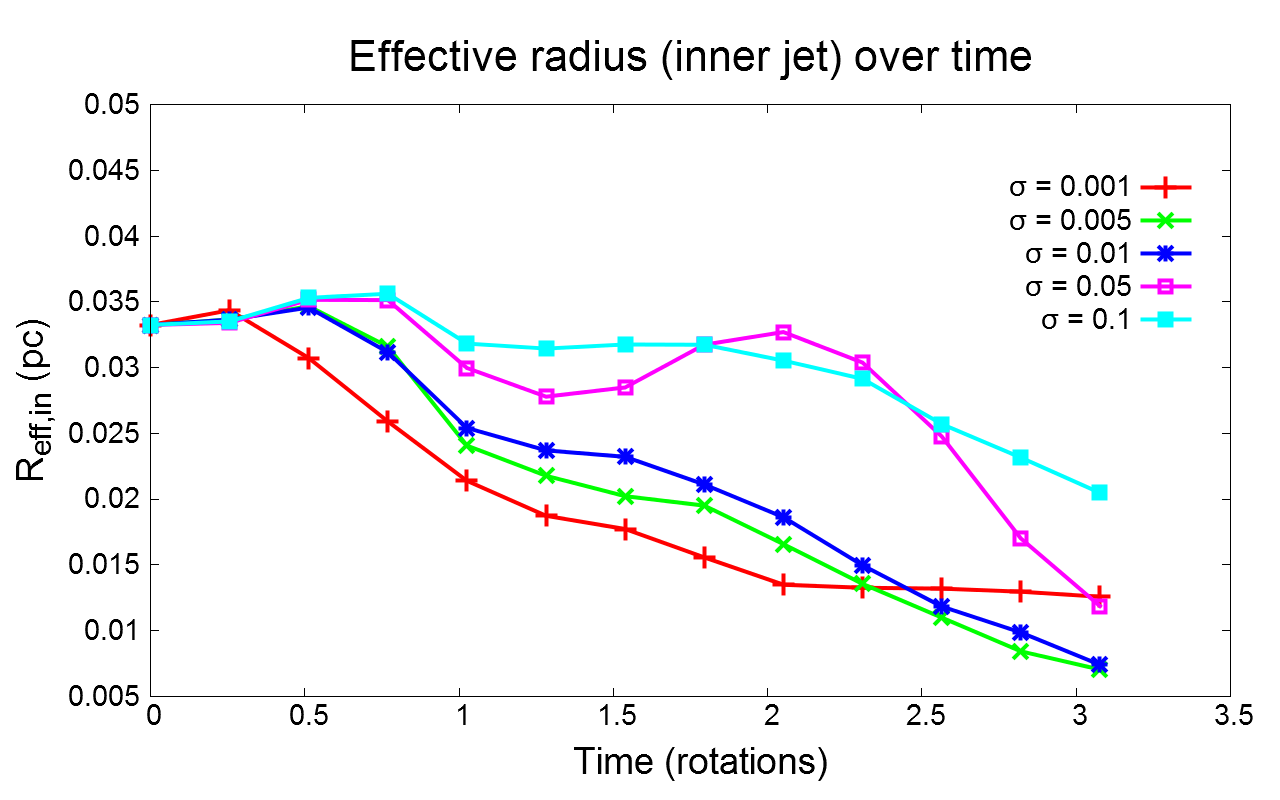}}	\hspace{1.5cm}
\subfloat[$\Gamma_{eff}$, $\sigma=0.001$, $t=3$rotations]
{\includegraphics[width=0.94\columnwidth, height=4.95cm]{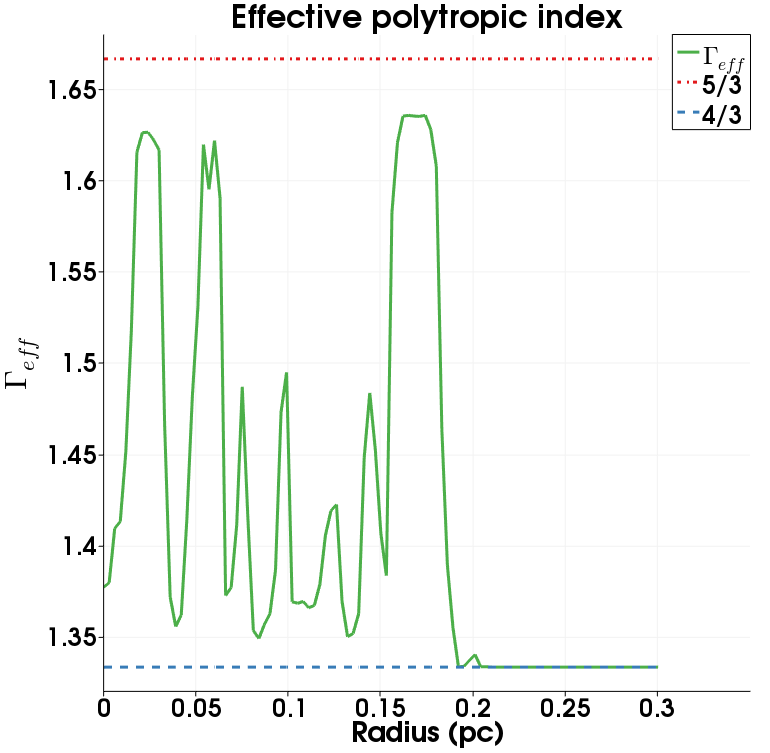}} 	

\subfloat[$R_{eff out}(t)$]
{\includegraphics[width=0.94\columnwidth, height=4.95cm]{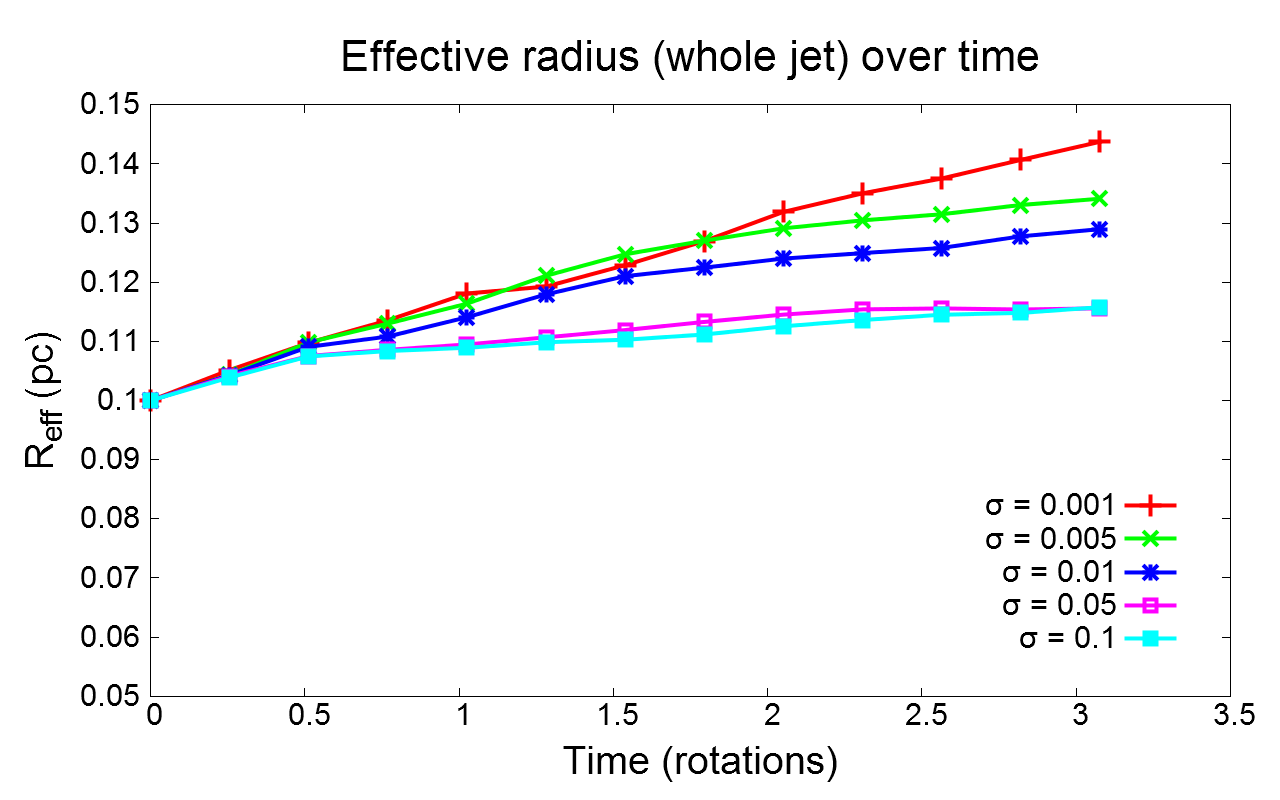}}  
 \hspace{1.25cm}
\subfloat[$\Gamma_{eff}$, $\sigma=0.01$, $t=3$rotations]
{\includegraphics[width=0.94\columnwidth, height=4.95cm]{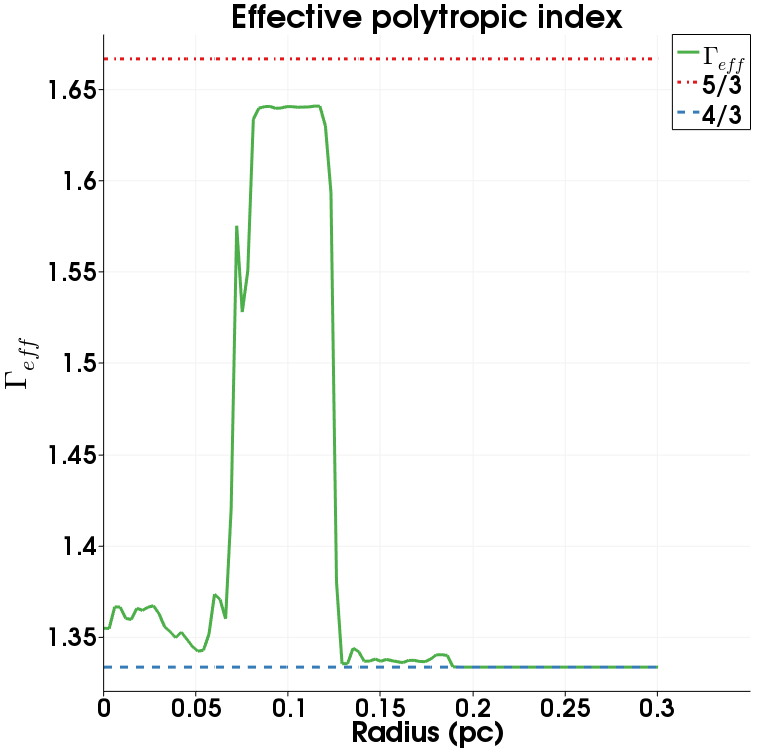}} 	

\subfloat[$\bar{\gamma}_{in}(\sigma)$]
{\includegraphics[width=0.94\columnwidth, height=4.95cm]{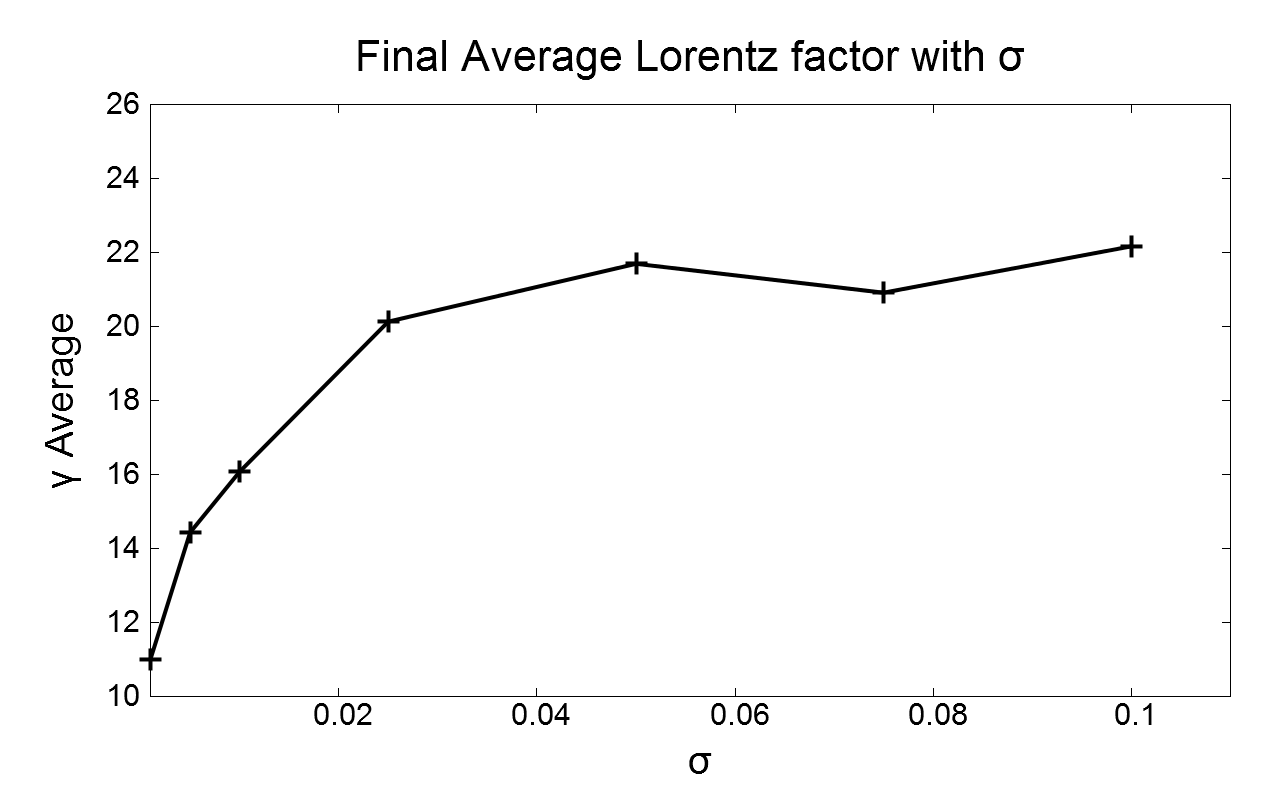}} \hspace{1.25cm}
\subfloat[$\Gamma_{eff}$, $\sigma=0.1$, $t=3$rotations]
{\includegraphics[width=0.94\columnwidth, height=4.95cm]{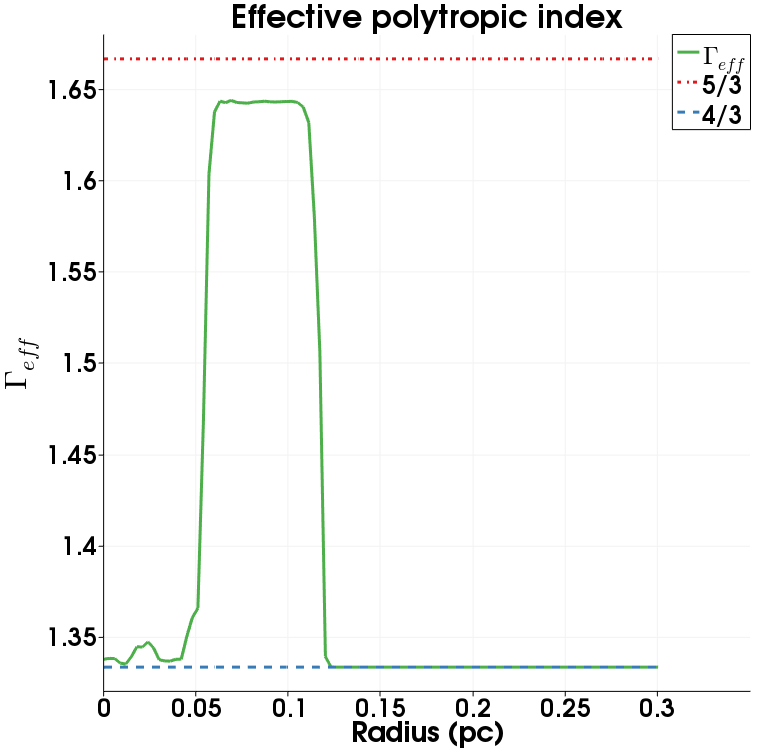}}  
\caption{Left column: (a) Average Lorentz factor of the inner jet, (c) Effective radius of the inner jet, (e) Effective radius of the outer jet after 3 rotations of the inner jet and (g) Final Average Lorentz factor with $\sigma$. Right column: (b) Initial effective polytropic index (for all cases) and Effective polytropic index after 3 rotations for (b) $\sigma=0.001$ (f) $\sigma=0.01$ (d) and $\sigma=0.1$ (h) \label{fig: trends1}}
\end{figure*}
Here we discuss the evolution of the average Lorentz factor and the effective radius of the jet, inner and outer, for cases with increasing magnetization, some of which were discussed in sections~\ref{sec: sig0_001}-~\ref{sec: sig0_1}. We also assess the effect of mixing in the effective polytropic index  and the density of each jet component.

In the absence of a toroidal magnetic field, it has been shown that the average Lorentz factor is significantly reduced in all unstable cases, while the effective radius of the jet (inner \& outer) increases \citep{Meliani2009}. This trend persisted, less obvious, also in more stable setups. We expect a similar behaviour in every case we examined here as well, with a less prominent decrease of the Lorentz factor at least in the very stable cases. The effective radius of each component is expected to increase in every case as well (as radial expansion is observed in every simulation), with the more stable cases resulting in a somewhat smaller increase (at least for the inner part).

First we define the criteria that we use to distinguish the two components. Specifically, we consider as inner jet the region in which the Lorentz factor has a value of $\gamma > 5$ and the density is $\rho<1$, whereas the whole jet is distinguished from the surrounding medium if $\gamma > 1.5$ and $B_z > 0.01$. We note however that the precise final result can be quite sensitive to the selection of the above criteria, especially when strong mixing is taking place. The value of the Lorentz factor is sufficient though to mask each region in all our cases so far.

For very low magnetization values, the average Lorentz factor is significantly reduced, as expected since we asymptotically approach the case of a purely poloidal magnetic field. We see a clear trend in the behaviour of  the average Lorentz factor of the whole jet with increasing magnetization, as seen in Fig.~\ref{fig: trends1}. As $\sigma$ becomes higher, the average Lorentz factor decreases at a slower rate and the effective radius of the whole jet also increases at a slower rate. For the most stable cases, the increase in the effective radius of the jet compared to the initial state, is of the order of 10\%.

In a similar way, we see a trend also in the evolution of the effective radius of the whole jet with $\sigma$. Increasing $\sigma$ results in a less significant expansion and a more structured outflow, for both components. The effective radius of the inner jet is affected by mixing, which slows down this part of the outflow. Due to the previous criteria, in cases where mixing is stronger (in other words, in the most unstable setups), a smaller region is identified as ``inner jet''. This deceleration is less prominent in stable cases. We notice a peculiar behaviour in the evolution of the case with $\sigma=0.05$ (Fig.~\ref{fig: trends1},second panel, left column). This is due to a transient state which results in an oval-like shape of the inner jet, present approximately from $\sim1.5-2.5$ rotations. 

Last, since the two jet components have different effective polytropic indices, it is interesting to check if the mixing has any effects in the final effective polytropic index of each region. For the unstable case (with $\sigma=0.001$), we notice that the effective polytropic index is significantly modified after 3 rotations. In the other two cases, with $\sigma=0.01$ and $\sigma=0.1$, we notice that the outer jet is more or less unaffected, as noted already in terms of density and the effective polytropic index is the same as in the initial state. The inner jet has a higher value of $\Gamma_{eff}$ compared to the initial state, but still closer to the relativistically hot value (especially for $\sigma=0.1$).
\section{Inner jet evolution at lower Lorentz factors}
In this section we examine the differences in the evolution of the jet if we a assume a lower Lorentz factor for the inner jet (closer to the values suggested by \cite{Giroletti2004}, \cite{PinerBG2003} for radio galaxies). We reduce the poloidal velocity of the inner jet so that $\gamma \simeq 10$ and we examine two values of magnetization, $\sigma = 0.001$ and $\sigma = 0.1$.

\subsection{Case 6: $\sigma=0.001$ \& $\gamma_{in}=10$}
This is a direct variant of Case 1, with a modified value for the Lorentz factor of the inner jet, $\gamma \simeq 10$.
\begin{figure}
\centering
\includegraphics[trim={0.5cm 0.7cm 0.5cm 0.5cm}, clip, scale=0.175]{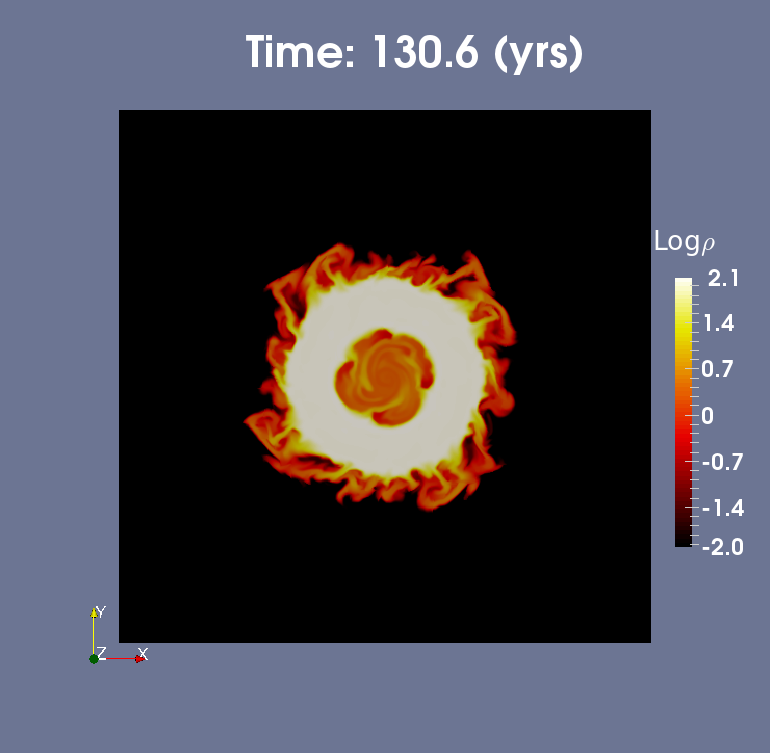}  \\
\includegraphics[trim={0.5cm 0.7cm 0.5cm 0.5cm}, clip, scale=0.175]{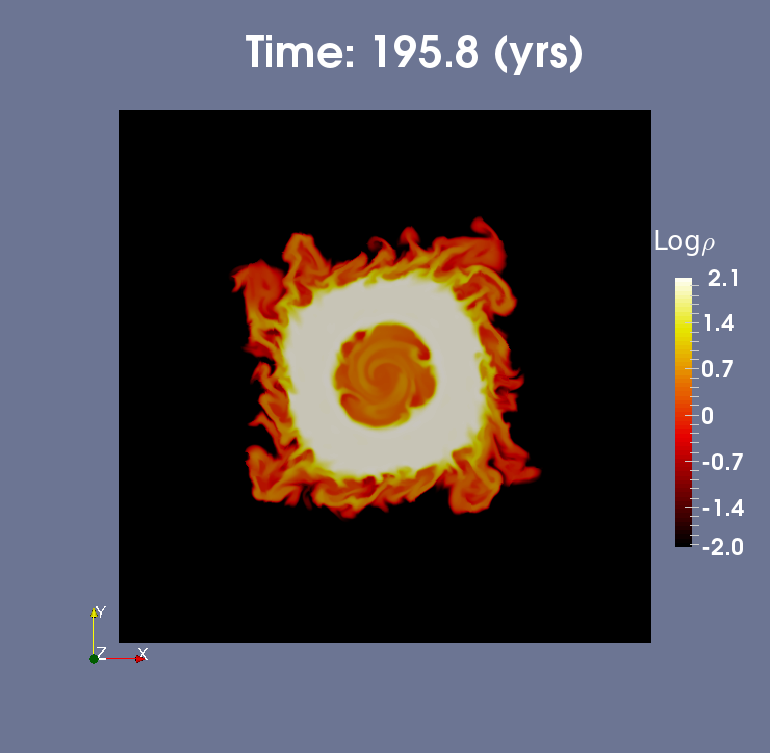}  \\
\caption{Proper density (log) for Case 6 (maximum $\sigma=0.001$). In this case the initial value of the Lorentz factor for the inner jet is 10. Snapshots after 2 and 3 rotations of the inner jet. \label{fig: lfac10sig0_001dens}} 
\end{figure}

Initially the evolution follows what has already been seen in the previous cases, with some notable differences. First, we can mostly distinguish the outer, high density part and the light, inner part of the jet throughout the evolution of the setup. Kelvin-Helmholtz instabilities can first be seen between the two jet components at a very early time (< 0.5 rotations of the inner jet and then vanishing), whereas Rayleigh-Taylor modes are mostly suppressed. In terms of Lorentz factor, both components remain structured and collimated after 3 rotations. The density and Lorentz factor distributions are shown in Fig.~\ref{fig: lfac10sig0_001dens} and Fig.~\ref{fig:  lfac10lfac} respectively.

Although Case 1 was the most unstable one, in terms of deceleration and decollimation, this case is considerably more stable, resembling in a sense the cases of higher magnetization of section~\ref{sec: results1}. The stability is related to the decrease of the inner jet inertia, since this is proportional to $\gamma^2$, as discussed later in the paper.
\subsection{Case 7: $\sigma=0.1$ \& $\gamma_{in}=10$}
In a similar fashion, we examine the case with highest magnetization (Case 5, $\sigma=0.1$) in a jet where we substitute again $\gamma_{z,in}\simeq10$.
\begin{figure}
\centering
\includegraphics[trim={0.5cm 0.7cm 0.5cm 0.5cm}, clip, scale=0.155]{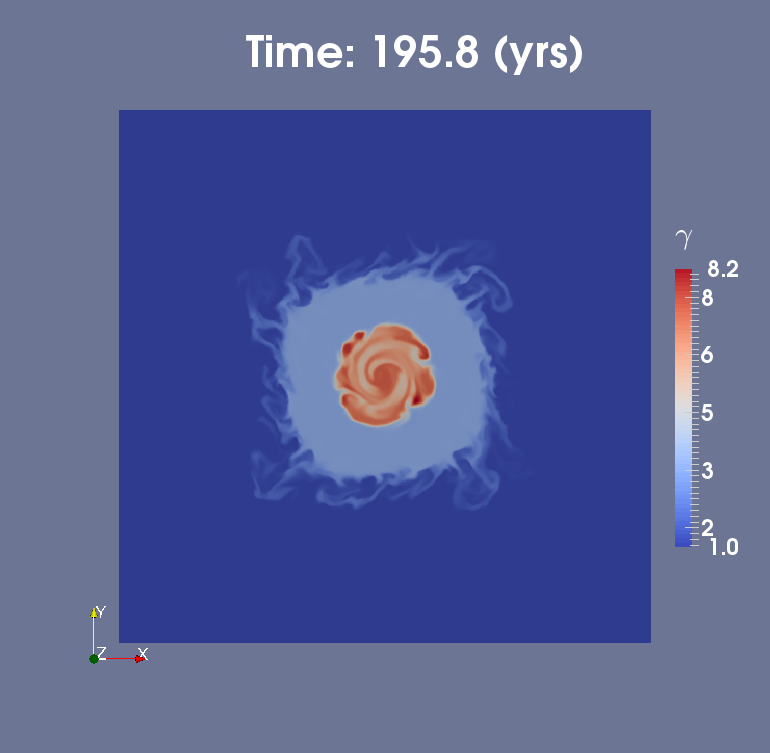}
{\includegraphics[trim={0.5cm 0.7cm 0.5cm 0.5cm}, clip, scale=0.155]{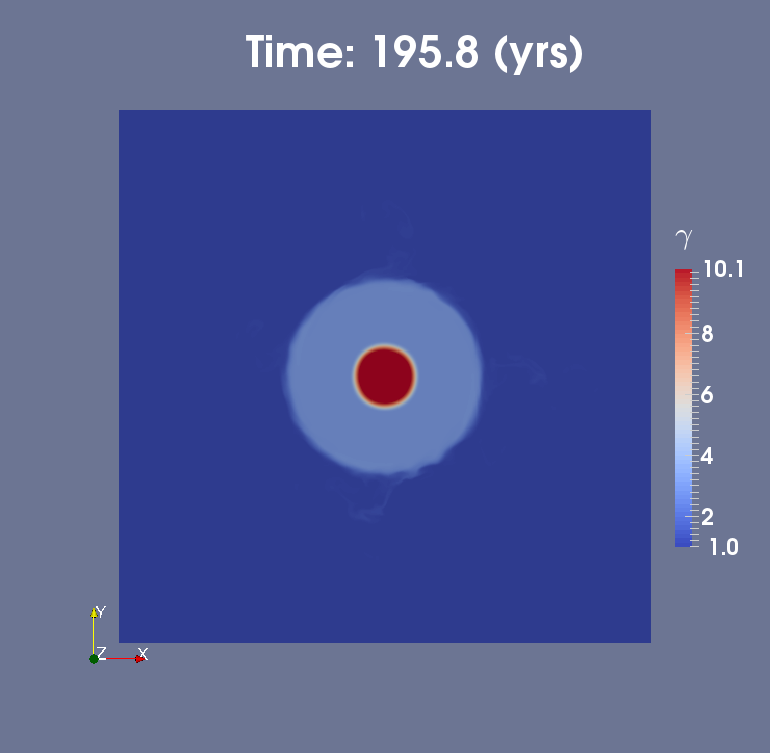}}
\caption{Lorentz factor for Cases 6 (left panel, maximum $\sigma=0.001$) and 7 (right panel, maximum $\sigma=0.1$) after 3 rotations of the inner jet. Both cases assume an initial Lorentz factor of 10 for the inner jet. We note that in both runs, the final state is much more stable compared to Cases 1 and 5. Especially for Case 7, the change, compared to the initial state, is negligible. \label{fig: lfac10lfac}}
\end{figure}

\begin{figure}
\centering
\includegraphics[trim={0.5cm 0.7cm 0.5cm 0.5cm}, clip, scale=0.175]{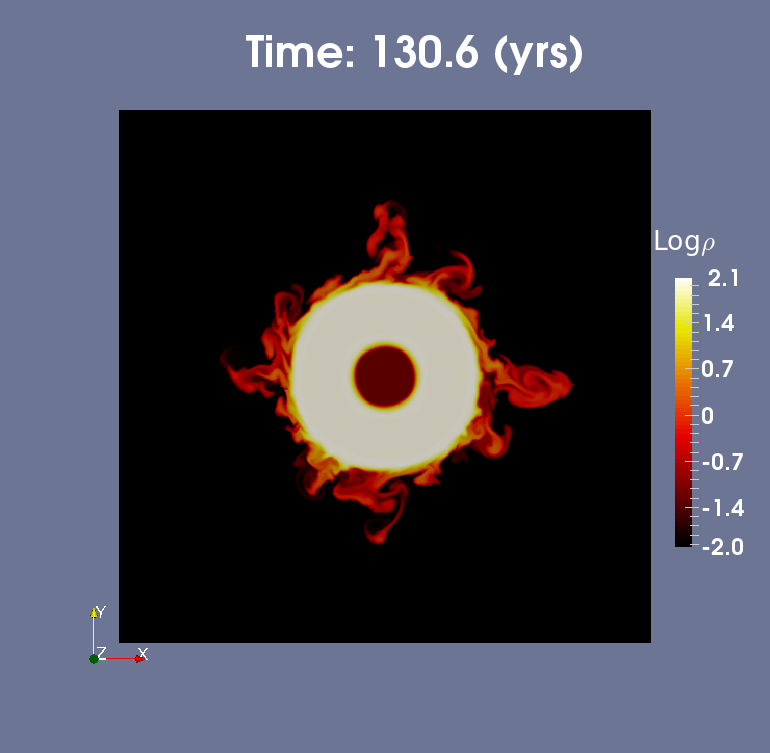}  \\
\includegraphics[trim={0.5cm 0.7cm 0.5cm 0.5cm}, clip, scale=0.175]{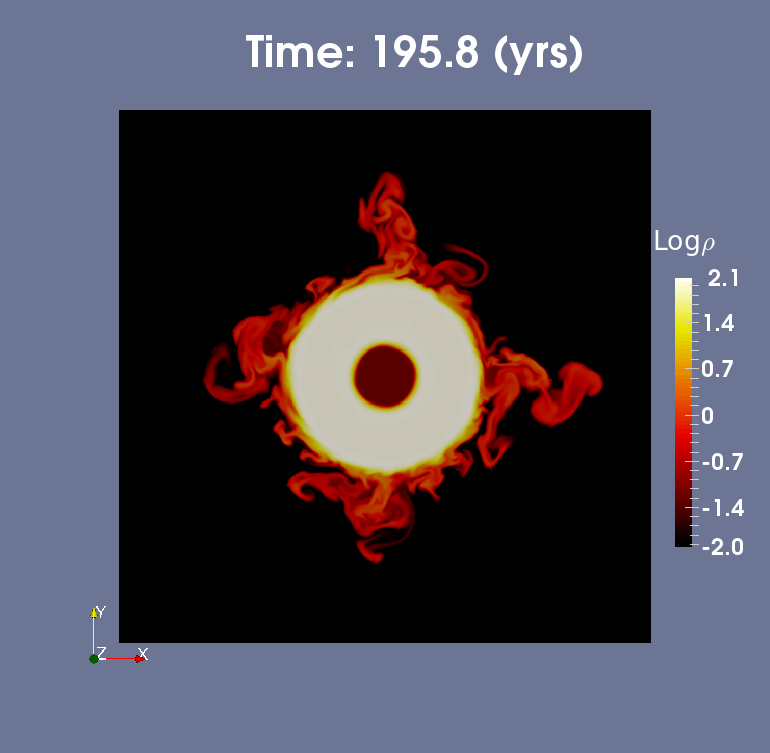}  \\
\caption{Proper density (log) for Case 7 (maximum $\sigma=0.1$). In this case the initial value of the Lorentz factor for the inner jet is 10. Snapshots after 2 and 3 rotations of the inner jet. \label{fig: lfac10sig0_1dens}} 
\end{figure}
Case 5 was the most stable, so we expect this feature to persist also here. Initially, there is little interaction between the two jet components.

Even at later times, the inner and outer parts of the jet experience little to almost no mixing and the two components can be clearly distinguished throughout the evolution of the jet, both in terms of Lorentz factor and density. The increased stability is again attributed to the lower effective inertia contrast between the components. For the density and Lorentz factor distributions, we refer to Fig.~\ref{fig: lfac10sig0_1dens} and Fig.~\ref{fig: lfac10lfac}.

We note that in both cases, lower inner jet inertia leads to more stable jets with respect to the Rayleigh-Taylor type instability, which is in agreement with the (hydrodynamic or purely poloidal magnetic field) results of \citet{Meliani2009}. This is not always the case though in other types of instabilities, e.g. Kelvin-Helmholtz, which are stabilized for higher values of jet velocity. This is shown in the case of a single component, matter dominated, cold jet in \citet{Bodo2013}.
\section{The non-rotating jet limit}
In this section we examine the evolution of the cases with $\sigma=0.001$ and $\sigma = 0.1$ (keeping $\gamma_{in} \simeq 30$) in the non-rotating limit, choosing a very slow value for $v_{\phi}$, namely $v_{\phi}= 10^{-6}$. This effectively cancels the centrifugal force and greatly stabilizes the outflow. Although in 2.5D we can not monitor Kelvin-Helmholtz instabilities along the z axis, which is dominant in non-rotating matter dominated jets according to \citet{Bodo2013}, we can check if the absence of rotation ceases to induce Rayleigh-Taylor type instabilities. 
\subsection{Case 8 \& 9: $\sigma = 0.001$ and $\sigma = 0.1$}
A notable difference in the non-rotating jet limit, is the absence of ``arms'' in the early stages of the evolution (up to $\sim 0.5$ rotations). 

Even for the lowest value of magnetization, $\sigma=0.001$, the centrifugal force is practically zero to play any role in the evolution of the outflow. A limited inward \& outward motion can be noticed between the two parts of the jet, which results in the formation of a small annulus of mixed material which surrounds the inner jet. 

In the case of $\sigma=0.1$ almost no mixing is observed. The two components remain separated throughout the run, clearly seen in the density plots. The average Lorentz factor at the end state is very similar to Case 7. This is an indication that in all cases where we assumed a rotating jet, the centrifugal force plays a key role in the mixing. 

\begin{figure}
\centering
\includegraphics[trim={0.5cm 0.7cm 0.5cm 0.5cm}, clip, scale=0.175]{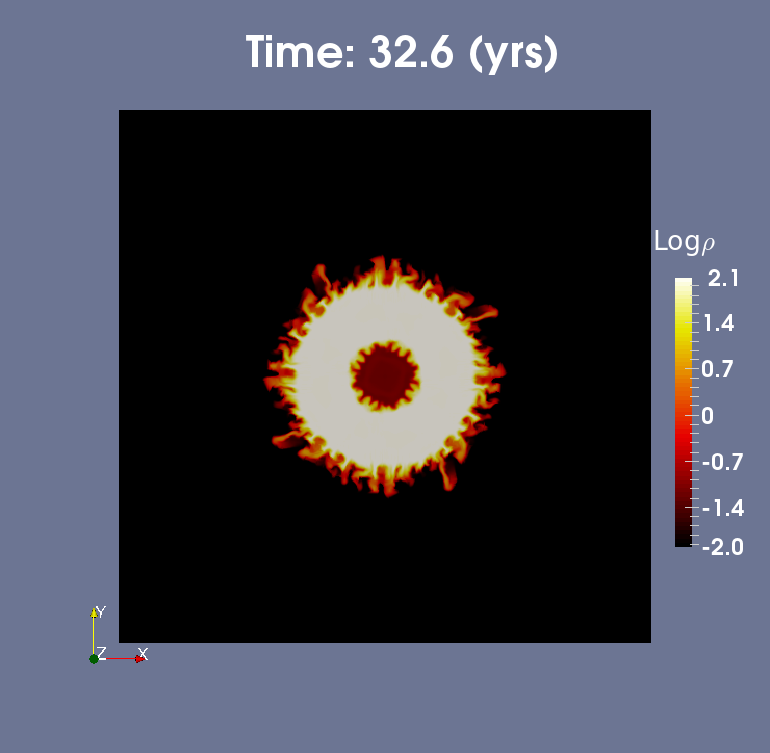} \\
\includegraphics[trim={0.5cm 0.7cm 0.5cm 0.5cm}, clip, scale=0.175]{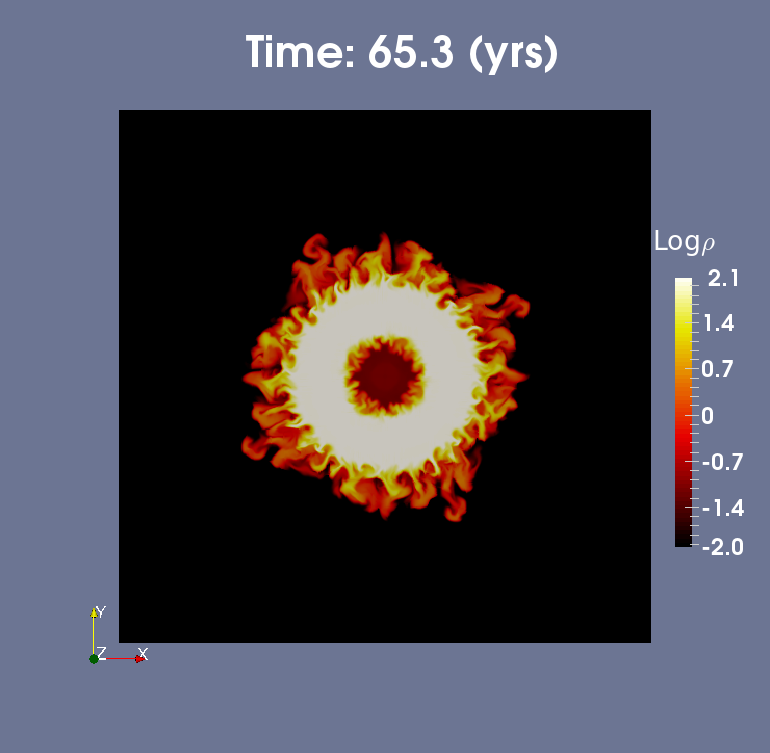}  \\
\includegraphics[trim={0.5cm 0.7cm 0.5cm 0.5cm}, clip, scale=0.175]{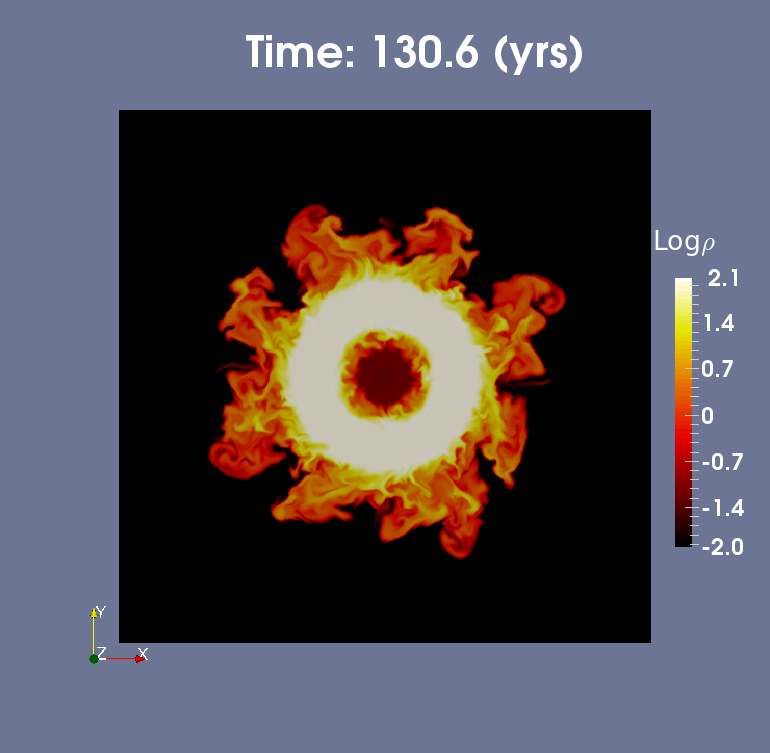}  \\
\includegraphics[trim={0.5cm 0.7cm 0.5cm 0.5cm}, clip, scale=0.175]{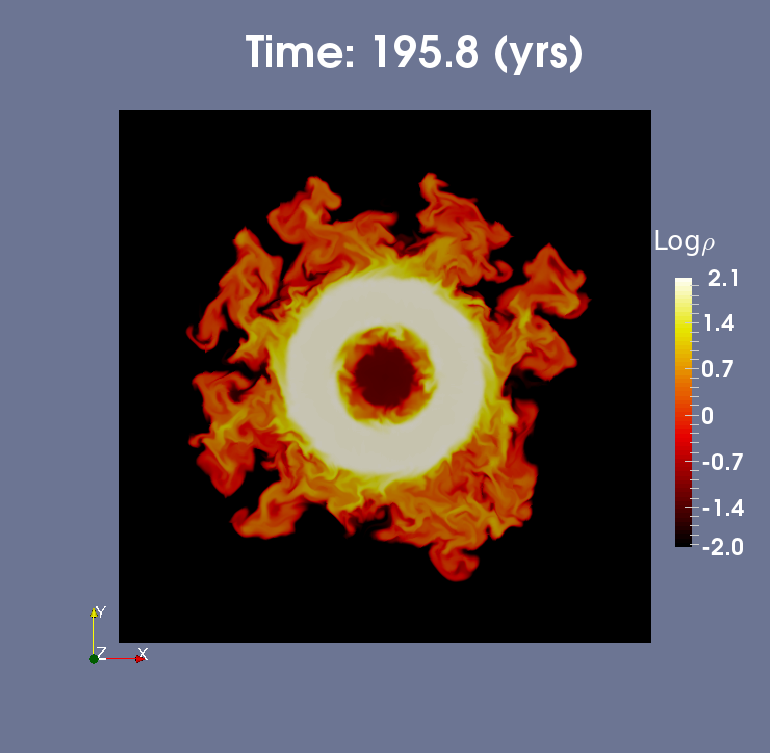}  \\
\caption{Proper density (log) for Case 8 (maximum $\sigma=0.001$). In this case the initial value of the rotation of the jet is close to zero. Snapshots after 0.5, 1, 2 and 3 rotations of the inner jet. The minimal centrifugal force leads to a stable final state compared to Case 1 \label{fig: nonrot1dens}} 
\end{figure}

\begin{figure}
\centering
\includegraphics[trim={0.5cm 0.7cm 0.5cm 0.5cm}, clip, scale=0.175]{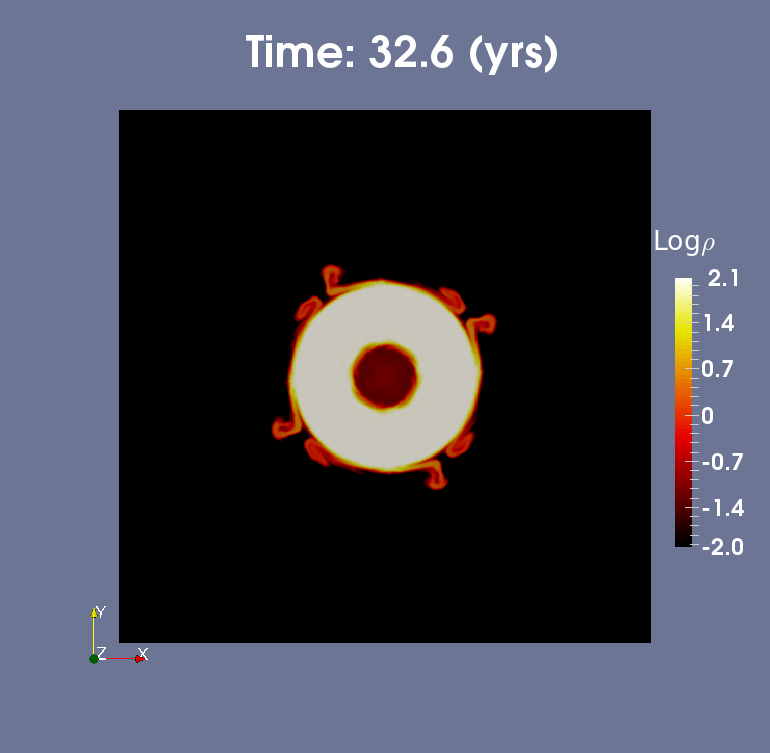} \\
\includegraphics[trim={0.5cm 0.7cm 0.5cm 0.5cm}, clip, scale=0.175]{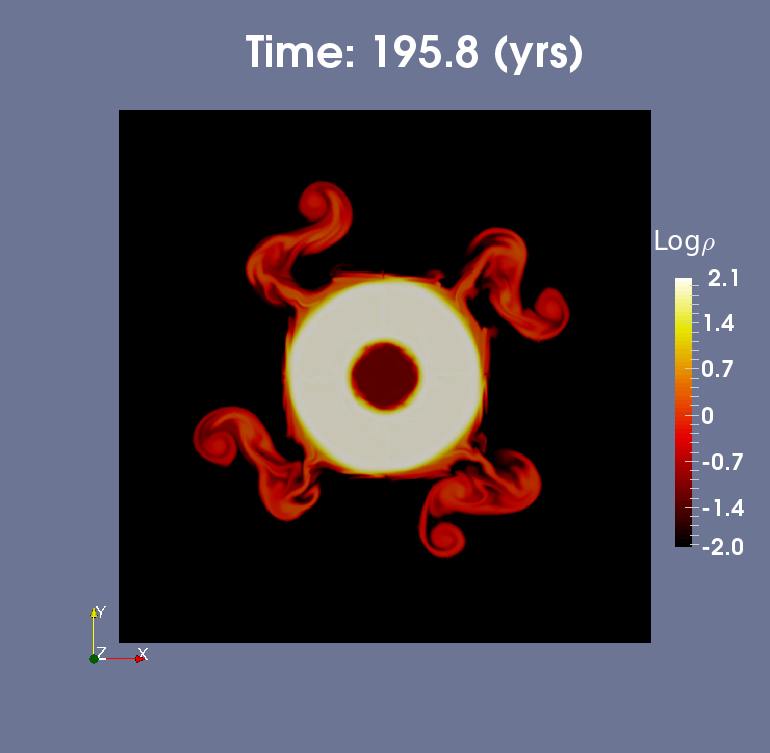}  \\
\caption{Proper density (log) for Case 9 (maximum $\sigma=0.1$). In this case the initial value of the rotation of the jet is close to zero. Snapshots after 0.5 and 3 rotations of the inner jet. The increased magnetic tension further stabilizes the outflow, compared with Case 8. \label{fig: nonrot2dens}} 
\end{figure}

The density distributions for each case are given in Fig.~\ref{fig: nonrot1dens} and Fig.~\ref{fig: nonrot2dens} respectively. In Fig.~\ref{fig: nonrot1dens}, we see that for $\sigma=0.001$, the outer jet still shows interchange/Rayleigh-Taylor type effects.
 
\subsection{Average Lorentz factor and Effective radius}
\begin{figure}
\centering
\subfloat[$\bar{\gamma}_{in}(t)$]
{\includegraphics[width=0.85\columnwidth, height=4.25cm]{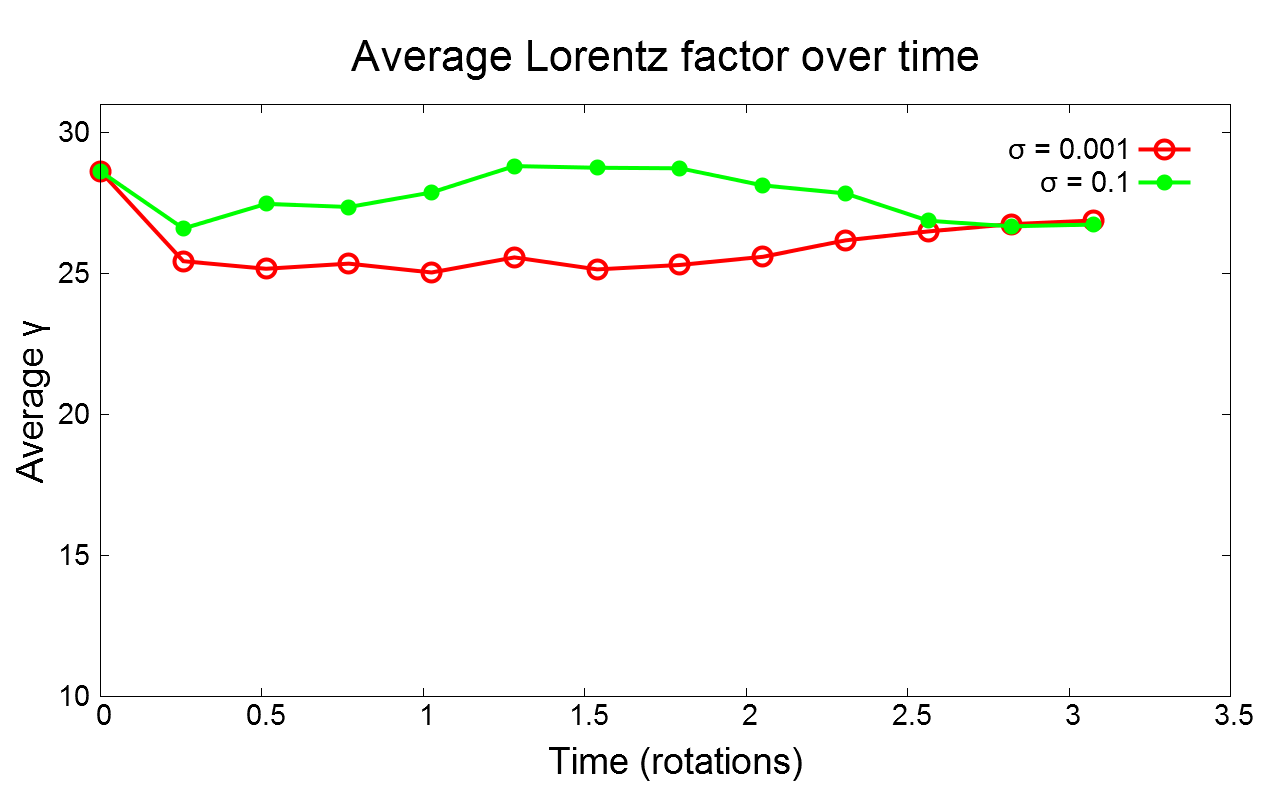}}  \\
\subfloat[$R_{eff,out}(t)$]
{\includegraphics[width=0.85\columnwidth, height=4.25cm]{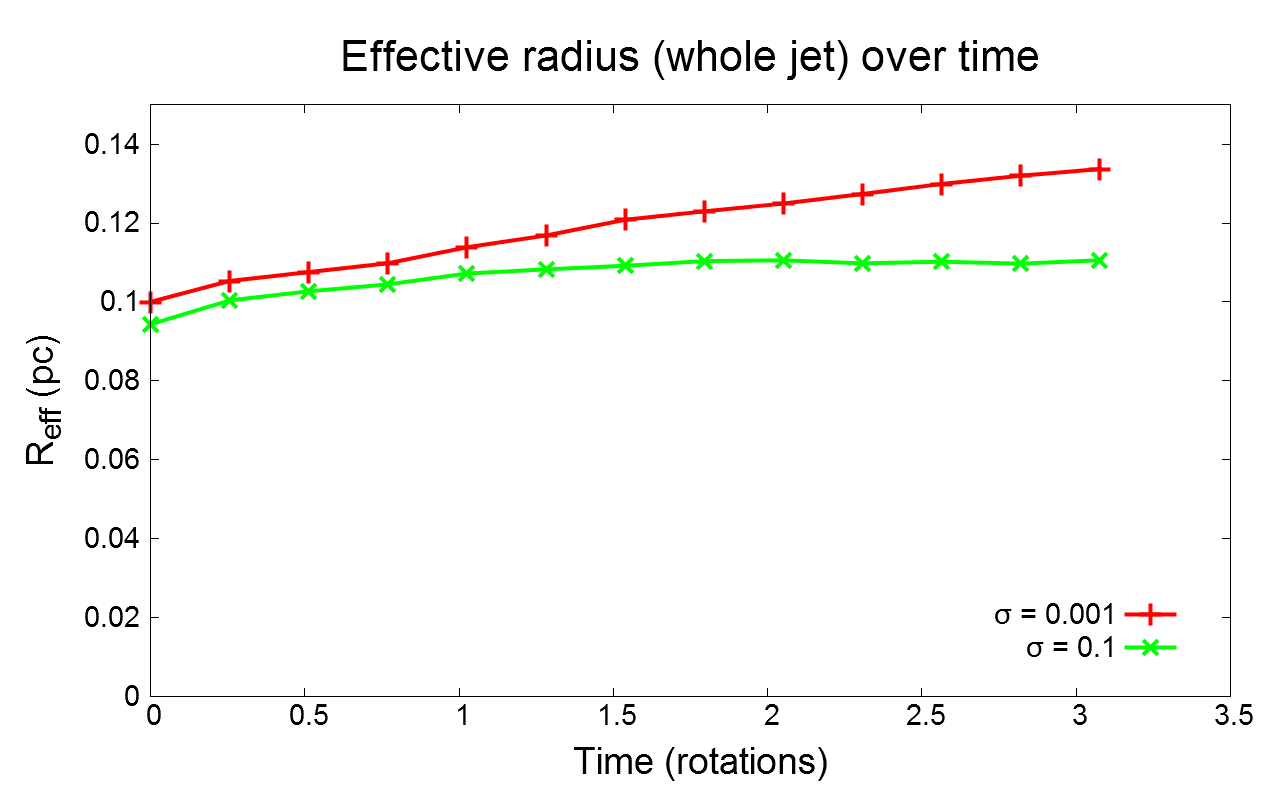}} 
\caption{Average Lorentz factor of the inner jet (top) and effective radius of the outer jet (bottom) after 3 rotations assuming a realistic Lorentz factor of 10 for the inner jet. \label{fig: Lfac10avg}}
\end{figure}

As stated before, for very stable cases we expect the average Lorentz factor to be more or less constant throughout the evolution of the jet. The same should hold for the effective radius, where we expect the outflow to expand less than in the unstable cases.

In the non-rotating limit, this holds, with a small increase in the effective radius of the inner jet due to the presence of a small shear region between the components. In principle, using more strict criteria (which may not be appropriate for the previous cases though), one can perform some fine tuning to clearly distinguish all different regions.

The effective radius of the outer jet (and thus the whole outflow) increases, due to the diffusion of dense material from the outer part to the static medium. This is more of interchange type, but does not affect significantly the integrity of the outer jet significantly. The trends (Lorentz factor \& effective radius over time for different $\sigma$) are presented in Fig.~\ref{fig: Lfac10avg}.

The increase in the effective radius of the whole jet is smaller compared to Cases 1-5. Even for $\sigma = 0.001$, the increase of the effective jet radius is $\sim 30$\% compared to $\sim 40$\% of Case 1. For a more detailed comparison, we refer again to Table \ref{tab: table_cases}.
\section{3D simulations}
In this section we examine some of the ``extreme'' scenarios from the previous sections, namely the 3D equivalent of Cases 1 ($\sigma=0.001$ and $\gamma_{in} \simeq 30$) and 7 ($\sigma=0.1$ and $\gamma_{in} \simeq 10$), which are the most unstable and the most stable cases respectively. In both cases we refer again to rotating jets, with $v_{\phi} = 0.01$ at $R=R_{in}$.

We remind that for the 3D simulations we use a uniform, Cartesian grid with a resolution of $256^3$. While this is lower than the effective resolution used in the 2.5D cases, it is still quite sufficient to capture small scale phenomena. The perturbation in the radial velocity, in both cases, satisfies $\nabla \cdot \vec{v} = 0$.
\subsection{Case 1b: $\sigma=0.001$, $\gamma_{in}=30$}
We examine the most unstable case of section ~\ref{sec: results1}, with $\sigma=0.001$. The aim is to check if the enhanced Rayleigh-Taylor type instability is still present (e.g. at $z=0.5pc$) and if other types of instabilities are dominant.

Initially, on the $x,y$ plane (see Fig.~\ref{fig: 3D_z05}) the evolution of the jet displays no major differences compared to the previous cases. Regarding the inner jet, since again no specific modes are excited, we notice the formation of 4 ``arms'', a feature which persist up to $\sim 1$ rotation. This is in agreement with the behaviour of all the low-$\sigma$ 2.5D setups. Along the $z$ axis, we notice the formation of Kelvin-Helmholtz instabilities around $\sim 0.3$ rotations, mainly between the outer jet and the external medium. 

A 3-slice representation of the computational domain is given in the left column of Fig.~\ref{fig: rho3D}. A view of the $x,y$ plane at $z=0.5pc$ is given in the top left (density) and top right (Lorentz factor) images in Fig.~\ref{fig: 3D_z05}. We highlight the rapid expansion of the outer jet due to the Kelvin-Helmholtz effects.

Overall, the setup is more unstable than the 2.5D case, in terms of expansion of the outer jet mainly due to the development of Kelvin-Helmholtz instabilities along the $z$ axis (but we note that causality can still play a role in decelerating the jet). Up to half rotation time, the two components can still be distinguished in terms of density and Lorentz factor, a trend which persists roughly up to a full rotation of the inner jet. Up to one rotation, there is no significant scaling of the density or the Lorentz factor with $z$, for the inner jet and a part of the outer jet up to $\sim 0.11pc$. This is shown in the bottom image of Fig.~\ref{fig: rho3D}, where we plot the density variation with $z$ in different regions (i.e. for different values of $R$). We use $R=R_{in}/2$ for the inner jet and $R=0.25pc$ for the interface between the outer jet and the external medium (where we notice the formation of vortices due to Kelvin-Helmholtz instabilities).
\subsection{Case 7b: $\sigma=0.1$, $\gamma_{in}=10$}
We examine one of the most stable cases, the 3D variant of Case 7, with $\sigma=0.1$ and $\gamma=10$ for the inner jet. Compared to the most unstable cases, the magnetic tension is now significant and the inertia reduced, thus we expect no significant Rayleigh-Taylor type instabilities. Kelvin-Helmholtz effects though can still be anticipated due to the density and velocity shear between the outer jet and the environment.

In the early stages of the simulation, up to $\sim 0.3$ rotations there is no significant change in the jet. After this time, Kelvin-Helmholtz instabilities begin to develop along the $z$ axis, between the outer jet and the external medium. In this case, the instabilities are somewhat less prominent due to the strong toroidal field compared to the previous case. We emphasize that the velocity shear between these media is the same in every case, since we keep the Lorentz factor of the outer jet fixed to $\gamma =3$.

\begin{figure}
\centering
\includegraphics[trim={0.75cm 0.7cm 0.5cm 1.6cm}, clip, scale=0.15]{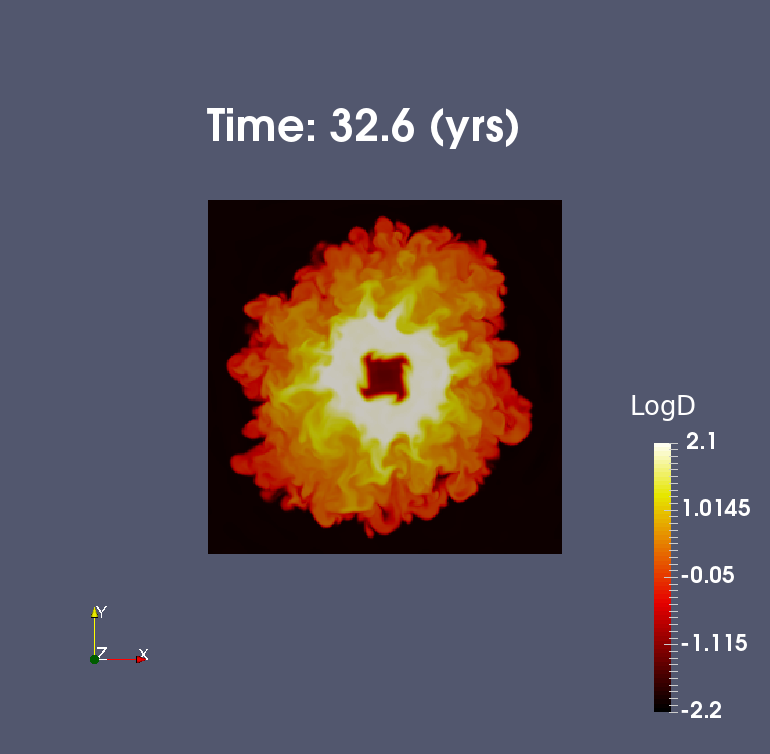}
\includegraphics[trim={0.75cm 0.7cm 0.5cm 1.6cm}, clip, scale=0.15]{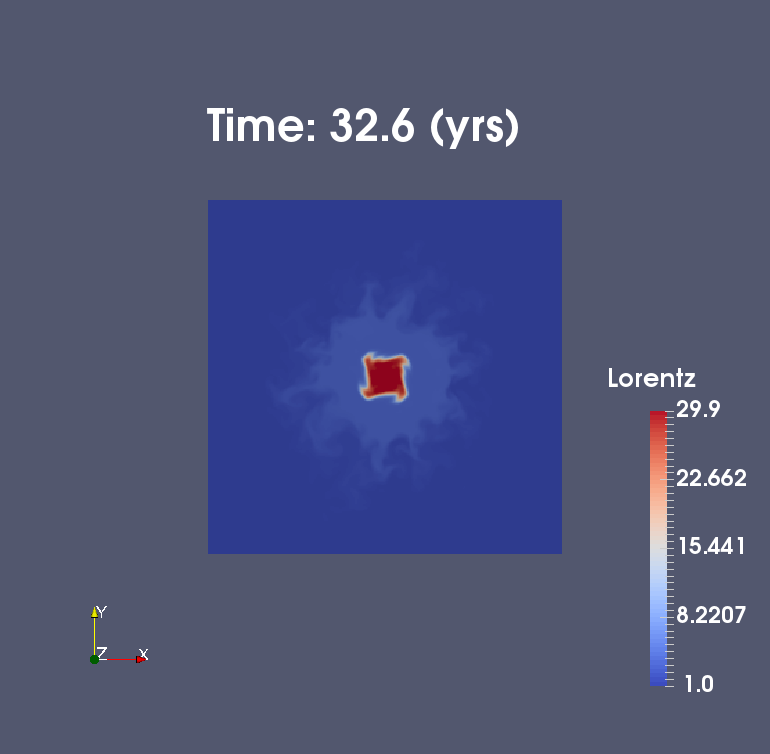}

\includegraphics[trim={0.75cm 0.7cm 0.5cm 1.6cm}, clip, scale=0.15]{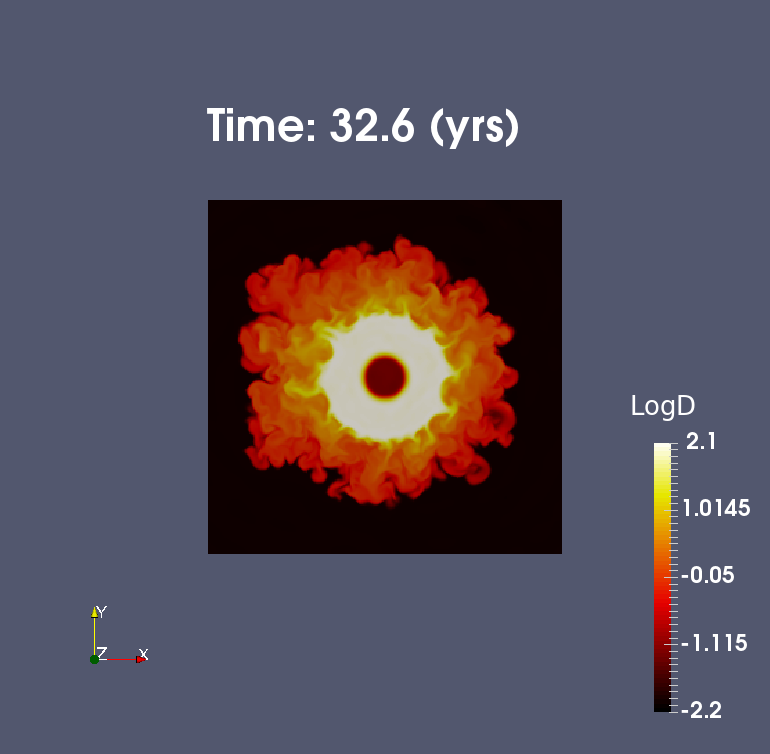}
\includegraphics[trim={0.75cm 0.7cm 0.5cm 1.6cm}, clip, scale=0.15]{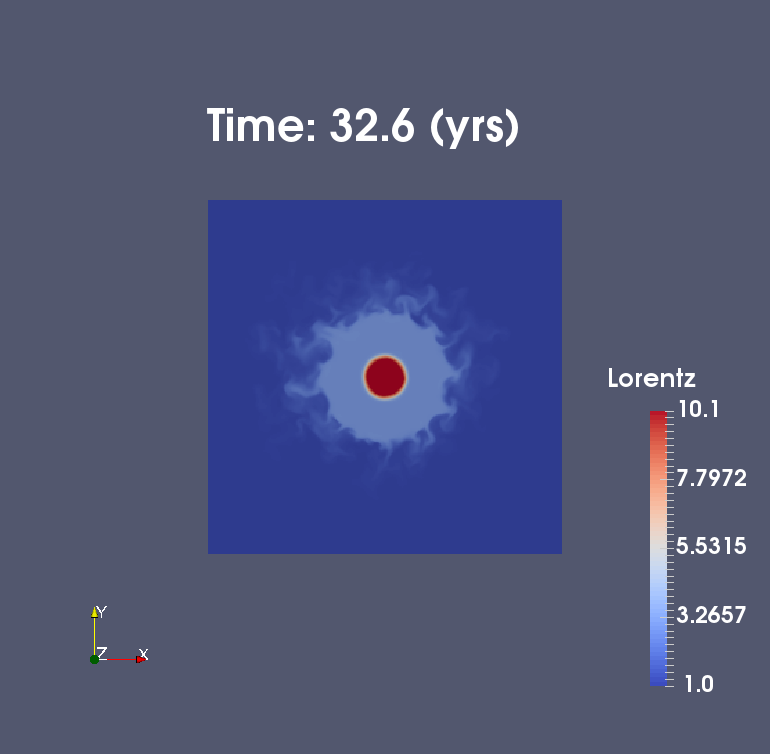}  
\caption{Left column: Proper density (log) for Cases 1b (top) and 7b (bottom), with maximum $\sigma=0.001$ and $\sigma=0.1$ respectively. Right column: same for the Lorentz factor. Snapshot after 0.5 rotations of the inner jet, view of the $x,y$ plane, at $z=0.5pc$. \label{fig: 3D_z05}} 
\end{figure}

\begin{figure}
\centering
\includegraphics[trim={0.23cm 0.7cm 0.5cm 0.2cm}, clip, scale=0.15]{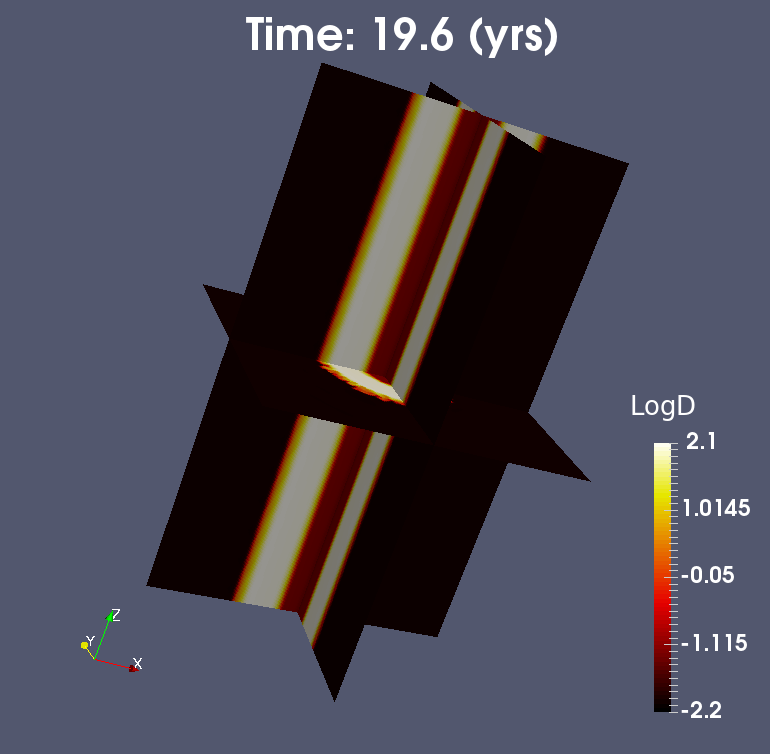}
\includegraphics[trim={0.23cm 0.7cm 0.5cm 0.2cm}, clip, scale=0.15]{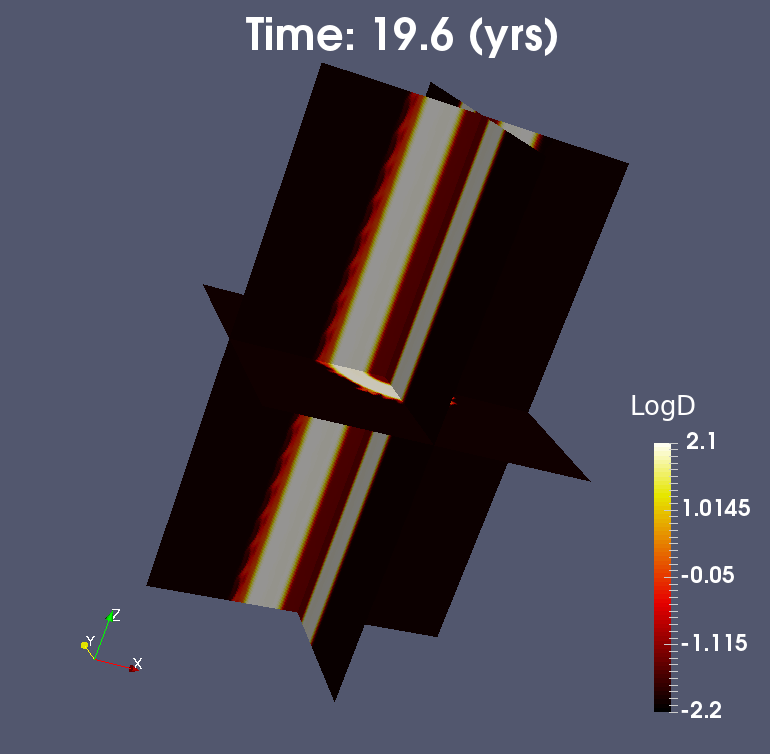}

\includegraphics[trim={0.23cm 0.7cm 0.5cm 0.2cm}, clip, scale=0.15]{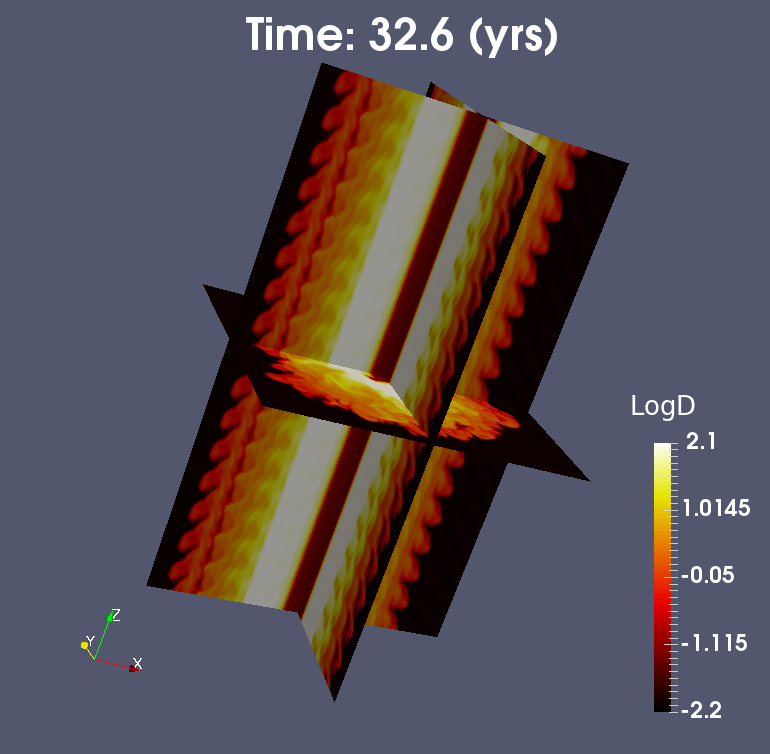}
\includegraphics[trim={0.23cm 0.7cm 0.5cm 0.2cm}, clip, scale=0.15]{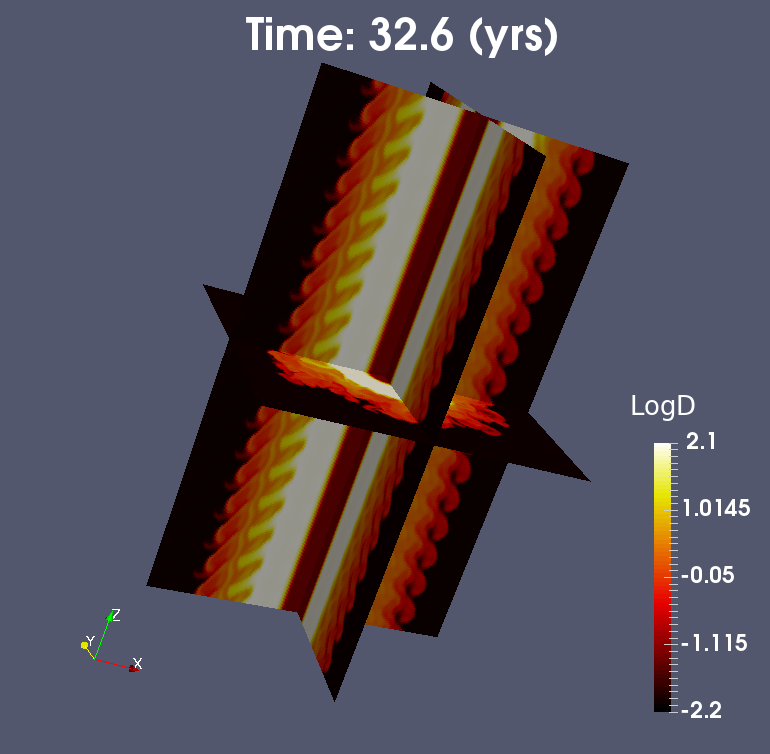} \vspace{0.15cm}

\includegraphics[trim={0.36cm 0.23cm 0.36cm 0cm}, clip, width=0.85\columnwidth, height=4.75cm]{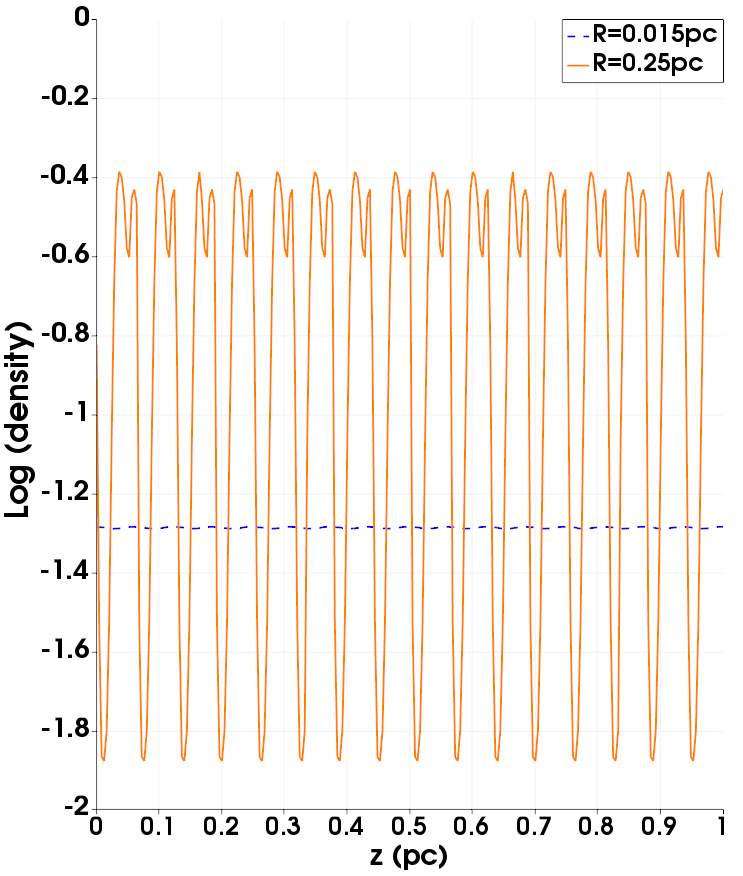}
\caption{Proper density for Cases 1b (left column, maximum $\sigma=0.001$, $\gamma=30$) and 7b (right column, maximum $\sigma=0.1$, $\gamma=10$). Snapshots after 0.3 and 0.5 rotations of the inner jet. The flow is moving upwards along the z axis. The bottom image shows the density variation along the $z$ direction, for different radii, after 0.5 rotations (Case 1b). The dashed line refers to $R=R_{in}/2$ and the solid line to the interface between the outer jet and the surrounding medium. The periodic behaviour of the density distribution with $z$ at the interface is an indication of Kelvin-Helmholtz instabilities. A similar behaviour is noticed in Case 7b.\label{fig: rho3D}} 
\end{figure}

A 3-slice representation can be seen in the right column of Fig.~\ref{fig: rho3D} and the $x,y$ plane at $z=0.5pc$ in Fig.~\ref{fig: 3D_z05}. We notice that the strong tension prevents the formation of ``arms'' between the inner and the outer jet and, at least in terms of Lorentz factor, the configuration is stable. 

Overall, the setup is similar to the relevant 2.5D case, with the two jet components remaining separate, distinguishable in terms of density and Lorentz factor, and the jet collimated, at least up to 1 rotation. In terms of expansion of the outer jet, we notice an increase in the effective radius of the outer part, again mainly due to the development of the Kelvin-Helmholtz instability along the $z$ axis. The outer jet has a reduced density, compared to the initial one, due to the mixing induced by Kelvin-Helmholtz. 

\section{Semi-analytical interpretation}
We will examine the above cases following a semi-analytical approach, similar to \citet{Meliani2009}. This interpretation focused on velocity perturbations in the radial direction only and excluded terms of $B_{\phi}$ in the momentum equation, e.g. magnetic tension. Since in our case the magnetic field includes a toroidal component, we want to examine the importance of the extra term(s) in the overall (in)stability of each different case. 

The momentum equation, using the term $\frac{d\vec{V}}{dt}$ is \citep{Kalra2000}:
\begin{gather}
\begin{split}
&\gamma^2 \Bigg[ \rho h +B^2 - \Big( \vec{V} \cdot \vec{B} \Big)^2 \Bigg] \frac{d \vec{V}}{dt} + \nabla \Big(p + \frac{B^2}{2} \Big) + \\
& \vec{V} \frac{\partial}{\partial t} \Big(p+ \frac{B^2}{2} \Big) - \Bigg\{ \Bigg(\frac{\vec{B}}{\gamma} \cdot \nabla \Bigg)\frac{\vec{B}}{\gamma} + \frac{\vec{B}}{\gamma} \Bigg(\nabla \cdot \frac{\vec{B}}{\gamma} \Bigg) + \\
& \Big(\vec{V} \cdot \vec{B} \Big) \Bigg[\frac{d\vec{B}}{dt} + \vec{B} \Big(\nabla \cdot \vec{V} \Big) + \Big(\vec{B} \cdot \nabla \Big) \vec{V} \Bigg] + \vec{B} \frac{d}{dt} \Big(\vec{V} \cdot \vec{B} \Big)  \Bigg\} = 0
\end{split}
\label{eq: mom_kalra}
\end{gather}
whereas the HD limit can be found in \citep{Bodo2004}. 

In \citet{Meliani2009}, the approximate momentum equation, ignoring charge separation, was:
\begin{equation} \label{eq: momentumold}
\Bigg( \gamma^2 \rho h + B_z^2 \Bigg) \Bigg[ \frac{\partial}{\partial t} + \vec{V} \cdot \nabla \Bigg] \vec{V} + \nabla p_{tot} + \vec{V} \frac{\partial p_{tot}}{\partial t} \simeq 0
\end{equation}
where the total pressure is $p_{tot} = p + \frac{B^2}{2}$. As in \citet{Meliani2009}, we evaluate the assumption of the minimal charge separation contribution (lab frame) in the momentum equation by examining the maximum radial and toroidal velocity components. In every case, $V_{\phi} < 2 \cdot 10^{-2}$ and $V_{r} < 10^{-3}$ and thus the contribution of the $\rho_e \vec{E}$ term is negligible.

Starting from equation~(\ref{eq: mom_kalra}), we extend this expression by including the toroidal component of the magnetic field while we maintain the assumption that the velocity is mainly poloidal ($v_z>>v_{\phi}$) and that $\gamma \simeq const.$ initially in both components. The momentum equation will then be:
\begin{equation} \label{eq: momentumnew}
\begin{split}
&\Bigg( \gamma^2 \rho h + B_z^2 +\gamma^2 B_{\phi}^2 \Bigg) \Bigg[ \frac{\partial}{\partial t} + \vec{V} \cdot \nabla \Bigg] \vec{V} + \nabla p_{tot} + \vec{V} \frac{\partial p_{tot}}{\partial t} \\
&-\frac{1}{\gamma^2} \Bigg(\vec{B} \cdot \nabla \Bigg) \vec{B} -2(\vec{V}\cdot \vec{B}) \Bigg(\vec{B} \cdot \nabla \Bigg) \vec{V}-\vec{B}\frac{d(\vec{V}\cdot \vec{B})}{dt} \simeq 0
\end{split}
\end{equation}
The importance of the toroidal magnetic field component is expressed initially in the first parenthesis and via the magnetic tension term of the Lorentz force, i.e. $(\vec{B} \cdot \nabla) \vec{B}$. In the first case, the ratio $\frac{\gamma^2 B_{\phi}^2}{\gamma^2 \rho h}$ is of the order $\frac{\gamma^2 \sigma}{h}$, which is less than 0.4 for $\sigma$ up to 0.1.

If we further approximate equation~(\ref{eq: momentumnew}) neglecting the time derivative of $p_{tot}$ and assume that the velocity perturbation does not depend on $z$ or $\phi$, we obtain:
\begin{equation}
\Bigg( \gamma^2 \rho h + B_z^2 +\gamma^2 B_{\phi}^2 \Bigg) \Bigg[ \frac{\partial v_R}{\partial t} - \frac{V_{\phi}^2}{R}\Bigg] + \frac{\partial p_{tot}}{\partial R} + \frac{B_{\phi}^2}{\gamma^2 R}  \simeq 0
\end{equation}
From this expression, we notice that the jet inertia increases as $\gamma^2$. Moreover, the increase of $B_{\phi}$ in the outer jet, further increases the jet stability. 

We expect the magnetic tension to balance the centrifugal force for sufficiently high values of magnetization. Indeed, the ratio of the centrifugal force to magnetic tension is $\frac{h}{\sigma}\gamma^2 V_{\phi}^2 \sim \frac{h}{\sigma}$. The inner part of the jet has a significantly high enthalpy, whereas the outer part is cold and the tension is dominant for higher values of $\sigma$. Moreover, simulations with lower values of $\gamma$ for the inner jet are also more stable, since the ratio is proportional to $\gamma^2$. 
\section{Conclusions}
We performed 2.5D and 3D simulations of relativistic, two component jets, assuming differential rotation and a toroidal magnetic field component of different magnitude in each case. The chosen parameters (e.g. the Lorentz factor $\gamma$) correspond to AGN jets and each (2.5D) simulation is restricted to a plane ($x,y$) perpendicular to the jet axis, far from the central source. In every case, we assumed that the jet has already been accelerated and collimated.

We keep the same density ratio and toroidal velocity profile (the latter with different coefficients) in each case, as described in section \ref{sec: setups}. Depending on the choice of the maximum magnetization, we result in a different final state, in terms of mixing between the components, the average Lorentz factor and the effective radius of the jet. 

The key difference in this work, in comparison with previous studies \citep{Meliani2007,Meliani2009}, is the addition of a toroidal component in the magnetic field, in agreement with observations of real astrophysical jets. This extra component results in an additional term in the momentum equation, which represents the magnetic tension. Although we also gradually (from lower to higher $\sigma$) arbitrarily decrease the contribution of the $B_{z,in}$ component of the magnetic field in the magnetic pressure, and thus the effective inertia of the inner jet, the big picture depends mainly on the choice of the magnetization. We also note that the assumption of a continuous toroidal velocity field leads to a stronger centrifugal force in the outer jet, in contrast with \citet{Meliani2009} where the toroidal velocity component was discontinuous.

In all cases of section ~\ref{sec: results1}, a shear region between the inner and the outer jet is formed, with mixing between the components also taking place. The shear region and the mixing are different in each setup, with cases of very low magnetization (below $\sigma = 0.01$) resulting in stronger mixing. Case 1 ($\sigma = 0.001$) is the most unstable, with the average Lorentz factor dropping to $\sim$ 11 and the jet expanding to an effective radius of $\simeq 0.09pc$. The overall image of the unstable cases, resembles that of \citet{Meliani2009}, where the jets were decelerated and expanding. 

For higher values of magnetization, we can still distinguish an inner region of high Lorentz factor, although the maximum value is reduced when compared to the initial one. This effect is prominent in Case 5 ($\sigma = 0.1$). Even though mixing is still present, its effect is not that strong, but still capable to decelerate the inner jet to $\gamma \sim 20$. In the end we find that a toroidal magnetic field, of sufficiently high $\sigma$, damps the Rayleigh-Taylor type instability and makes the interface between the inner and the outer jet stable.

Furthermore, we examine cases with slower inner jets, to achieve a more realistic value of Lorentz factor $\gamma \sim 10$, which tend to be more stable even at low magnetizations. This is attributed to the lower effective inertia of the inner jet. If we keep the rotation of the jet to a minimum, the centrifugal force is negligible, unable to trigger any Rayleigh-Taylor type instabilities and thus the outflow still remains collimated after 3 rotations. These cases of course can still be unstable to other types of instabilities, e.g. Kelvin-Helmholtz or kink, which require a 3D study and a relaxation of the translational symmetry along the $z$ axis.

As regards the 3D simulations, at least for the asymptotic cases we examined, jets with weak toroidal magnetic field tend to remain unstable to the Rayleigh-Taylor type instability. Increasing the magnetization, we achieve stability but we note that both cases are still unstable to Kelvin-Helmholtz instabilities, which is dominant now up to 1 rotation. In this study we are modelling but a small region of the jet in 3D, with the upper $z$ boundary being equal to 10 jet radii. Using periodic boundary conditions though in the $z$ direction, we can represent a larger region (of sub-parsec to parsec scale) of the jet.

Overall, our results are complementary to the findings of \citet{Meliani2009}, where the proposed models were used as an analogy to explain the FRI/FRII classification. Furthermore, the (in)stability of such an outflow has impact on its structure (e.g. in terms of velocity, mixing etc.), which in turn can affect the emission we observe from these objects. This work will be continued in a follow-up study, synthetic observations will be created based on our 2.5D \& 3D results. With the translational symmetry relaxed, it is also possible to see if different kinds of instabilities (e.g. kink) develop. Furthermore, we plan to simulate different kinds of outflows, e.g. GRB jets and create synthetic radiation maps, where polarized emission can also be considered.

\section*{Acknowledgements}
This research was supported by projects GOA/2015-014 (2014-2018 KU Leuven) and the Interuniversity Attraction Poles Programme by the Belgian Science Policy Office (IAP P7/08 CHARM). The computational resources and services used in this work were provided by the VSC (Flemish Supercomputer Center), funded by the Research Foundation Flanders (FWO) and the Flemish Government - department EWI. Visualization was performed using Paraview (more information on www.paraview.org) and Gnuplot (http://www.gnuplot.info/index.html). DM thanks Bart Ripperda, Norbert Magyar, Fabio Bacchini \& Jo Raes for their comments on various stages of this work and Chun Xia \& Oliver Porth for their suggestions regarding the use of the code. Finally, we thank the anonymous referee for the useful comments on the manuscript.
 


\bibliographystyle{mnras}
\bibliography{references_Millas_et_al_MNRAS2017} 


\appendix

\section{Integration constants in pressure equation}

The full expressions for $\zeta$ in equation~(\ref{eq: preseq}) are \citep{Meliani2009}:
\begin{gather}
\begin{split}
& \zeta_{in} = p_o + \frac{\Gamma_{in}-1}{\Gamma_{in}}\rho_{in} + \frac{(\Gamma-1)(\tilde{a}+2)}{2\tilde{a}(\alpha(\Gamma-1)+\Gamma)}(-b_{\phi}^2+(b_{\phi}V_z-v_{\phi}B_z)^2) 
\end{split}
\end{gather}
for the inner jet and 
\begin{gather}
\begin{split}
& \zeta_{out} = \Bigg[ P_{tot}(R_{in}) + \frac{\Gamma_{out}-1}{\Gamma_{out}}\rho_{out} - \frac{B_{z,out}^2}{2} \Bigg]  \Big(1-\tilde{a}_{out}\Big)^{\frac{\Gamma_{out}}{\alpha_{out}(\Gamma_{out}-1)}}  \\
& + \Bigg[ -b_{\phi,out}^2+\Big(b_{\phi,out}V_{z,out}-v_{\phi,out}B_{z,out}\Big)^2 \Bigg] \\
& \times \Big(1-\tilde{a}_{out}\Big)^{\frac{\Gamma_{out}}{\alpha_{out}(\Gamma_{out}-1)}} \\
& \times \Bigg( \frac{1}{2}+\frac{(1-\tilde{a}_{out})(\Gamma_{out}-1)(\alpha_{out}+2)}{2\tilde{a}_{out}[\alpha_{out}(\Gamma_{out}-1)+\Gamma_{out}]} \Bigg)
\end{split}
\end{gather}
for the outer jet. We remind that $\Gamma_{in}= 4/3$ and $\Gamma_{out}=5/3$. 

\bsp	
\label{lastpage}
\end{document}